\documentclass[12pt]{article}
\usepackage{geometry}             
\usepackage{graphicx}
\usepackage{amssymb}
\usepackage{amsmath}
\usepackage{epstopdf}
\usepackage{comment}
\usepackage{cite}
\usepackage{abstract}

\usepackage[hyperindex=true,
          pdfstartview=FitH,
          bookmarksnumbered=true,
          bookmarksopen=true,
          citecolor=blue,
          linkcolor=blue,
          colorlinks=true,
          unicode]{hyperref}

\begin{document}
\title{Maxwell's equal area law for Lovelock Thermodynamics}
\author{Hao Xu and Zhen-Ming Xu\\
School of Physics, Nankai University,
Tianjin 300071, China\\
{\em email}:
\href{mailto:haoxu@mail.nankai.edu.cn}{haoxu@mail.nankai.edu.cn},
\href{mailto:xuzhenm@mail.nankai.edu.cn}{xuzhenm@mail.nankai.edu.cn}}

\author{Hao Xu \thanks{{\em
        email}: \href{mailto:haoxu@mail.nankai.edu.cn}
        {haoxu@mail.nankai.edu.cn}},
        Zhen-Ming Xu \thanks{{\em
        email}: \href{mailto:xuzhenm@mail.nankai.edu.cn}
        {xuzhenm@mail.nankai.edu.cn}}\\
School of Physics, Nankai University, Tianjin 300071, China\\
}

\date{}                             
\maketitle

\begin{abstract}
We present the construction of Maxwell's equal area law for the Guass-Bonnet AdS black holes in $d=5,6$ and third order Lovelock AdS black holes in $d=7,8$. The equal area law can be used to find the number and location of the points of intersection in the plots of Gibbs free energy, so that we can get the thermodynamically preferred solution which corresponds to the first order phase transition. We obtain the radius of the small and large black holes in the phase transition which share the same Gibbs free energy. The case with two critical points is explored in much more details. The latent heat is also studied.
\end{abstract}

\section{Introduction}
The geometric description of the black hole is described by Einstein's
General relativity. However, since the groundbreaking work of Hawking and
Bekenstein, new features began to show. People found black holes have thermodynamic properties. In the presence of a negative cosmological constant, more thermodynamic aspects emerge. The black holes possess rich phase structure and admit critical behavior.
The phase transition between stable large black holes and thermal gas in Schwarzschild AdS space-time was found in 1980s, which is known as Hawking-Page phase transition\cite{HawkingPage:1983}. It has been explained by Witten as the gravitational dual of the QCD confinement/deconfinement transition\cite{Witten:1998zw,Witten:1998qj}. Another important example is a first order phase transition which is reminiscent of the van der Waals liquid-gas transition in Reissner-Nordstrom AdS(RN AdS) spacetime\cite{ChamblinEtal:1999a,ChamblinEtal:1999b,Liu:2014gvf}.

Recently, this analogy is extended to more general cases \cite{KastorEtal:2009,Dolan:2010,Dolan:2011a,Dolan:2011b}. By identifying the negative cosmological constant as a thermodynamical variable, we can reconsider the critical behavior of AdS black holes in an extended phase space, i.e. the $P-V$ diagram. The black hole mass $M$ should be regarded as the enthalpy $H$ rather than the internal energy \cite{KastorEtal:2009}. We interpret the cosmological constant $\Lambda$ as an effective pressure, while its conjugate variable $V$, which is defined as $V=\frac{\partial M}{\partial P}$, can be taken as thermodynamical volume. The physical meaning of the thermodynamical volume remains to be fully understood, but it is conjectured to satisfy the reverse isoperimetric inequality\cite{CveticEtal:2011}, which indicates that the spherical boundary minimizes surface area for a given volume. However, recent works find some black holes violate this inequality, such as the rotating AdS black hole in ultraspinning limit \cite{Hennigar:2014cfa} and Lifshitz black holes \cite{Brenna:2015pqa}. These black holes are called super-entropic.

Many papers have been pursuing the above ideas for different choices of AdS black holes with positive, zero, or negative constant curvature \cite{D.Kubiznak,Poshteh:2013pba,Belhaj:2013cva,Altamirano:2013uqa,Altamirano:2013ane,Altamirano:2014tva,Wei:2012ui,
Cai:2013qga,Zou:2014mha,Chen:2013ce,Hristov:2013sya,Belhaj:2013ioa,Hendi:2012um,
Gunasekaran:2012dq,Zou:2013owa,Ma:2013aqa,Mo:2014qsa,Frassino:2014pha,Johnson:2014yja,Johnson:2014xza,Johnson:2014pwa,Xu:2014kwa,
Dolan:2014vba,Lee:2014tma,Zhang:2014uoa,Wei:2014qwa,Zhang:2015ova,Wei:2015ana,Rajagopal:2014ewa,Hennigar:2015esa,Frassino:2015oca,Oltean:2015lta,Mo:2015xpa,Lee:2015wua}.
In our previous work\cite{Xu:2013zea} we investigate the critical behavior of Gauss-Bonnet AdS black holes by taking the Gauss-Bonnet coupling constant as a free variable\cite{Kastor:2010gq}. In another work \cite{Xu:2014tja}, we studied the criticality of the third order Lovelock black holes. We find a single critical point for spherical horizon in $d=7$, and two for $d=8,9,10,11$. There is always a single critical point in $d\geq7$ for hyperbolic horizon.

In RN AdS space-time, there is an oscillating part at low enough temperature on $P-V$ plane. It predicts a first order phase transition in the plot of Gibbs free energy. This oscillating part should be replaced by an isobar which satisfies the Maxwell's equal area law and represents the small-large black holes coexistence line. The phase transition is characterized by the Hawking radiation with a constant temperature. One may wonder whether the pressure can vary during the phase transition or not. In fact, the isobar is a constant temperature line in constant pressure, so the AdS background geometry remains the same in the phase transition\cite{Spallucci:2013osa}. There have been some papers studying the this idea for different AdS black holes\cite{Spallucci:2013osa,Spallucci:2013jja,Belhaj:2014eha,Zhang:2014fsa,
Zhao:2014eja,Lan:2015bia}. The equal area law construction can be equally applied on the $T-S$ plane. There are also some suggestions that the holographic entanglement entropy (HEE) may also undergo van der Waals-like phase transition \cite{Johnson:2013dka,Caceres:2015vsa,Nguyen:2015wfa,Sun:2016til} on $T$-HEE plane. However, up to now the dual field theory interpretation of this phase transition remains to be an open problem.

When calculating the integral of $PdV$, different authors make different choices for the volume. Some people use the thermodynamic volume $V=(\frac{\partial H}{\partial P})_S$, while others use the specific volume $v$, which is proportional to the black hole radius. Both of them can obtain the correct critical exponents. However, we want to state that only the thermodynamic volume $V$ should be used in the equal area law. During the phase transition the Gibbs free energy is unchanged. On the other hand, the integral of $Pdv$ do not have the scaling of Gibbs free energy, so the resulting solutions of the black hole do not share the same Gibbs free energy. The equal area law cannot be used on the $P-v$ or $T-v$ plane\cite{Lan:2015bia}.

Motivated by the above consideration and progress, we extended the study of Maxwell's equal area law to Gauss-Bonnet gravity in $d=5,6$ and the third order Lovelock gravity in $d=7,8$. This is not a trivial repetition of the previous approach to new models. The authors of \cite{Belhaj:2014eha} investigated the charged Gauss-Bonnet AdS black holes, but they used the specific volume $v$ in the equal area law. In this work we apply the thermodynamic volume $V=(\frac{\partial H}{\partial P})_S$. We will show the small and large black holes indeed have the same Gibbs free energy. For the third order Lovelock black holes in
$d=8,9,10,11$, there can be very rich phase structures in the extended phase space, such as the reentrant phase transition which includes both first and zeroth order transition \cite{Xu:2014tja,Altamirano:2013ane,Gunasekaran:2012dq}. There exist two critical points $T_{c1}$ and $T_{c2}$. The equal area law for this kind of system has not been studied before. Isothermal plots of the Gibbs free energy in $T_{c1} < T< T_{c2}$ do not always predict the existence of thermodynamically preferred coexistence state of small and large black holes. We show this is because the equal area law corresponding to the first order phase transition is not always satisfied in the region $T_{c1}<T<T_{c2}$. There is a value $T_t$ located between $T_{c1}$ and $T_{c2}$. In the region $T_t<T<T_{c2}$, rather than $T_{c1}<T<T_{c2}$, we can find the first order phase transition. For $T<T_t$ only one phase of large black holes exists. In this paper, we use the numerical method to give the value of $T_t$ in $d=8$. $T_t$ is the "real critical temperature" which can predict the swallow tail corresponding to the thermodynamically preferred coexistence state of small and large black holes. Furthermore, we could get the latent heat in the first order phase transition.

The paper is organized as follows. In the next section, we provide a brief review about thermodynamics of Guass-Bonnet AdS black holes. Then we study the equal area law in $d=5,6$. In Section 3 we investigate the third order Lovelock black hols in $d=7,8$. Finally in Section 4 we present some concluding remarks.

\section{Equal area law for Gauss-Bonnet AdS Black holes}
We start with a brief review of the thermodynamics of Gauss-Bonnet AdS Black holes. The detailed calculation can be found in \cite{Cai:2013qga,RGCai2002}.
Setting the Newton constant $G=1$, the action is written as
\begin{align}
{\cal I}=\frac1{16\pi}\int d^dx \sqrt{-g}[R-2\Lambda+\alpha_{GB} (R_{\mu\nu\gamma\delta}R^{\mu\nu\gamma\delta}-4R_{\mu\nu}R^{\mu\nu}+R^2)],
\end{align}
where $\alpha_{GB}$ is the Gauss-Bonnet coefficient, or known as the second-order Lovelock
coefficient and cosmological constant $\Lambda=-\frac{(d-1)(d-2)}{2l^2}$. The
Lovelock coefficients are proportional to the inverse string tension
with positive coefficients in string theory, so in this paper we only consider the case $\alpha_{GB}>0$.

The static solution of the black hole takes the form\cite{Boulware,RGCai2002,Wiltshir,Cvetic,Kofinas:2006hr,Kastor:2011qp,Cai:2013qga}
\begin{align}
ds^2=-f(r)\mathrm{d}t^2+\frac{1}{f(r)}\mathrm{d}r^2+r^2 \mathrm{d}
\Omega_k^2,\label{eq:2a}
\end{align}
where
\begin{align}
f(r)=k+\frac{r^2}{2\alpha}\Bigg(1-\sqrt{1+\frac{64\pi\alpha M}{(d-2)\Sigma_k r^{d-1}}-\frac{64\pi\alpha P}{(d-1)(d-2)}}\Bigg),
\end{align}
$k=0,\pm1$ and $d\Omega_k^2$ is the line element of a $(d-2)$-dimensional
maximally symmetric Einstein manifold with curvature $(d-2)(d-3)k$ and volume $\Sigma_k$. $M$ is the black hole mass, $P=-\frac{\Lambda}{8\pi}$ and $\alpha=(d-3)(d-4)\alpha_{GB}$. The radius of the
AdS black hole $r_+$ is the largest root of $f(r)=0$. By identifying $H\equiv M$, we can obtain the enthalpy and temperature with the help of two equations $f(r_+)=0$ and $T=\frac{f'(r_+)}{4\pi}$, which yield
\begin{equation}
H\equiv M=\frac{(d-2)\Sigma_k r_+ ^{d-3}}{16\pi}\bigg(k+\frac{k^2\alpha}{r_+^2}+\frac{16\pi P r_+^2}{(d-1)(d-2)}\bigg),
\end{equation}
\begin{equation}
T=\frac{16\pi P r_+ ^4/(d-2)+(d-3)k r_+ ^2+(d-5)k^2 \alpha}{4\pi r_+  (r_+ ^2+2k \alpha)}.
\label{T1}
\end{equation}

It can be observed $d=5$ is different from other dimensions, since the factor $(d-5)$ controls the lowest power term of $r_+$ in Eq.\eqref{T1}. It is perhaps $d=2k+1$ is the lowest dimension in which the k-th order Lovelock density can affect
the local geometry. We can find the same phenomenon in the third order Lovelock black holes, where $d=7$ is different
from the others. The critical behavior in these dimensions will also be distinguished from the higher dimensions.

We can calculate the other thermodynamic quantities. The black hole entropy\cite{RGCai2002}
\begin{equation}
S=\frac{\Sigma_k r^{d-2}_+}{4}\bigg(1+\frac{2(d-2)\alpha k}{(d-4)r_+^2}\bigg),
\label{S}
\end{equation}
and the thermodynamic volume\cite{Cai:2013qga}
\begin{equation}
V=\bigg(\frac{\partial H}{\partial P}\bigg)_{S,\alpha}=\frac{\Sigma_k r_+ ^{d-1}}{d-1}.
\label{vol}
\end{equation}

The thermodynamic volume is a monotonic function of the horizon
radius. It is easy to check these thermodynamic quantities satisfy the first law of black
hole thermodynamics in the extended phase space
\begin{align}
\mathrm{d}H=T\mathrm{d}S+V\mathrm{d}P+\psi \mathrm{d}\alpha.
\label{firstlaw}
\end{align}
We treat the Gauss-Bonnet coefficient as a variable here and $\psi$ is the thermodynamic conjugate of $\alpha$\cite{Kastor:2010gq}.
In addition, the Gibbs free energy can be obtained by Legendre transformations as
\begin{equation}
G=G(T,P)=H-TS.
\label{Gibbs1}
\end{equation}

Eq.\eqref{T1} can be rearranged into the following form,
\begin{equation}
P=\frac{d-2}{4r_+}\left(1+\frac{2k\alpha}{r_+^2}\right)T-\frac{(d-2)(d-3)k}{16\pi r_+^2}-\frac{(d-2)(d-5)\alpha k^2}{16\pi r_+^4}.
\label{P}
\end{equation}

Now we consider Maxwell's equal area law. When the temperature is low enough, there may exist three black holes with different sizes at the same pressure. The medium sized black hole is unstable because its heat capacity is negative. The oscillating part of the isotherm should be replaced by an isobar which implies the small black hole jumps to a large one. This is a first order phase transition. The location of the isobar is determined by the Maxwell's equal area law since the Gibbs free energy remains unchanged during the phase transition. It satisfies following two equations \cite{Spallucci:2013osa}
\begin{equation}
\begin{aligned}
& P(r_1,T)=P(r_2,T)\\
& P(r_2,T) (V_2-V_1)=\int_{r_1}^{r_2}P(r_+,T)\mathrm{d}V
\label{law}
\end{aligned}
\end{equation}
where $P=P(r_1,T)=P(r_2,T)$ is an isobar defining equal areas, $r_1$ and $r_2$ are the radius of the small and large black holes respectively. When we get the solution $(r_1,r_2)$, the latent heat $L$ between the two phases can be easily obtained as
\begin{equation}
L=T\big(S(r_2)-S(r_1)\big)=\frac{\Sigma_k T}{4}\Big( r_2^{d-2}-r_1^{d-2}+\frac{2(d-2)\alpha k}{d-4}(r_2^{d-4}-r_1^{d-4})\Big),
\label{latent}
\end{equation}
which measures the loss(gain) of the black hole mass during the phase transition.

We consider the case of $d=5$ firstly. Inserting \eqref{P} and \eqref{vol} into \eqref{law}, we can have the solutions shown as following two cases.
\subsubsection*{1) The case of $k=0$}
The equal area law \eqref{law} becomes
  \begin{equation}
\begin{aligned}
& \frac{3T}{4r_1}=\frac{3T}{4r_2} \\
&\frac{3T}{4r_2}\Big(\frac{r_2^4}{4}-\frac{r_1^4}{4}\Big)=\int_{r_1}^{r_2}\frac{3T}{4}r_+^2\mathrm{d}r_+,
\label{lawa}
\end{aligned}
\end{equation}
  the only solution is
     \begin{equation*}
     r_1=r_2=const
     \end{equation*}
     which is obviously trivial. There is only one branch of locally thermodynamically stable black holes. The Eq.\eqref{P} is identical to the equation of state of the ideal gas. There is no critical behavior.

\subsubsection*{2) The case of $k\neq 0$}

When $k\neq 0$, \eqref{law} takes the form
  \begin{equation}
  \begin{aligned}
&\frac{3}{4r_1}\Big(1+\frac{2k\alpha}{r_1^2}\Big)T-\frac{3k}{8\pi r_1^2}=\frac{3}{4r_2}\Big(1+\frac{2k\alpha}{r_2^2}\Big)T-\frac{3k}{8\pi r_2^2}, \\
&\Big(\frac{3}{4r_2}\Big(1+\frac{2k\alpha}{r_2^2}\Big)T-\frac{3k}{8\pi r_2^2}\Big) \Big(\frac{r_2^4}{4}-\frac{r_1^4}{4}\Big)=\int_{r_1}^{r_2}\Big(\frac{3T}{4}r_+^2-\frac{3kr_+}{8\pi}+\frac{3k\alpha T}{2} \Big)\mathrm{d}r_+.
\label{lawb}
\end{aligned}
\end{equation}
Rearranging above two equations, the first one becomes
\begin{equation}
2\pi Tr_1^2 r_2^2=k r_1 r_2(r_1+r_2)-4\pi k \alpha T(r_1^2+r_1r_2+r_2^2),
\label{law1}
\end{equation}
and the second one is
\begin{equation}
2\pi T(r_1^2+r_1r_2+r_2^2)=3k(r_1+r_2)-36\pi k \alpha T.
\label{law2}
\end{equation}
Combining \eqref{law1} and \eqref{law2} together, we have
\begin{equation}
r_1 r_2(r_1+r_2)=4\pi \alpha T(r_1^2+4r_1 r_2 +r_2^2)
\label{law3}
\end{equation}
Inserting \eqref{law3} into \eqref{law1} we obtain
\begin{equation}
r_1r_2=6k\alpha.
\end{equation}
We introduce a new coefficient $y$ which satisfies
\begin{equation}
     r_1=y \sqrt{6 \alpha k}, \quad r_2=\frac{\sqrt{6 \alpha k}}{y},
     \label{solution1}
     \end{equation}
 then inserting \eqref{solution1} into \eqref{law3} we have
   \begin{equation}
    \label{quartic}
    4 \pi \sqrt{\alpha} T y^4-\sqrt{6 k}y^3+16\pi \sqrt{\alpha}T y^2-\sqrt{6 k}y+4\pi\sqrt{\alpha}T=0.
   \end{equation}
 Since $\alpha >0$, $k$ must be non-negative, so we will take $k=+1$. When $y=1$, $r_1=r_2=\sqrt{6 \alpha}$. It is exactly the critical radius obtained in \cite{Cai:2013qga}. The corresponding temperature $T=\frac{1}{\pi\sqrt{24\alpha}}$, which is the critical temperature $T_c$. In this case, the size of the three black holes becomes identical. The phase transition is second order. When $T>T_c$, no real root of \eqref{quartic} can be found, so there is no critical behavior. When $T<T_c$, we can find one solution $(r_1,r_2)$. This is also the location where the first order phase transition happens.

It is natural to consider the limit $T\rightarrow 0$. We can take the limit $T\rightarrow 0$, but $T$ can not equal to zero because $T$ controls the lowest power term of $r_+$ in $P$, which is $3\alpha T/2r_+^3$. If $T=0$, there will be no oscillating part so that the equations of equal area law \eqref{law} and the solution \eqref{solution1} and \eqref{quartic} are not valid anymore. From solution \eqref {quartic} we know
\begin{equation}
T=\frac{\sqrt{6}(y^3+y)}{4\pi \sqrt{\alpha}(y^4+4y^2+1)}.
\end{equation}
If $T\rightarrow 0$, we have $y\rightarrow 0$ or $y\rightarrow +\infty$ from the above formula. Since $r_2=\frac{\sqrt{6 \alpha k}}{y}\geq y \sqrt{6 \alpha k}\geq 0$, $y$ must satisfy $0\leq y\leq 1$, so we will take $y\rightarrow 0$, which leads to $r_2\rightarrow +\infty$ and $r_1\rightarrow 0$. To the leading order $T\sim \frac{\sqrt{6}y}{4\pi \sqrt{\alpha}}=\frac{3}{2\pi r_2}$.
We could also insert $T\sim \frac{3}{2\pi r_2}$, $r_2\rightarrow +\infty$ and $r_1\rightarrow 0$ to the second condition of Eq. \eqref{law}. The LHS is
\begin{eqnarray}
\Big[\frac{3}{4r_2}\Big(1+\frac{2\alpha}{r_2^2}\Big)T-\frac{3}{8\pi r_2^2}\Big] \Big(\frac{r_2^4}{4}-\frac{r_1^4}{4}\Big)
&=&\frac{3}{16}T r_2^3-\frac{3}{32\pi}r_2^2+\frac{3\alpha}{8}Tr_2\\ &\sim& \frac{3}{16\pi}r_2^2,
\end{eqnarray}
while the RHS is
\begin{eqnarray}
\int_{r_1}^{r_2}\Big[\frac{3}{4r_+}\Big(1+\frac{2\alpha}{r_+^2}\Big)T-
\frac{3}{8\pi r_+^2}\Big]\mathrm{d}\Big(\frac{r_+^4}{4}\Big)&=&\frac{1}{4}Tr_2^3-\frac{3}{16\pi}r_2^2+\frac{3\alpha}{2}Tr_2\\
&\sim&\frac{3}{16\pi}r_2^2.
\end{eqnarray} The LHS and RHS are equal in $T\rightarrow 0$. The term $ Tr_2^3\sim r_2^2$. From the \eqref{latent} we obtain the latent heat \begin{equation}
L=T\big(S(r_2)-S(r_1)\big)=\frac{\Sigma_k T}{4}\big(r_2^3-r_1^3+6\alpha k(r_2-r_1)\big).
\end{equation}
When $T\rightarrow 0$, $L\sim Tr_2^3\sim r_2^2\rightarrow +\infty$.

\begin{figure}
\begin{center}
\includegraphics[width=0.35\textwidth]{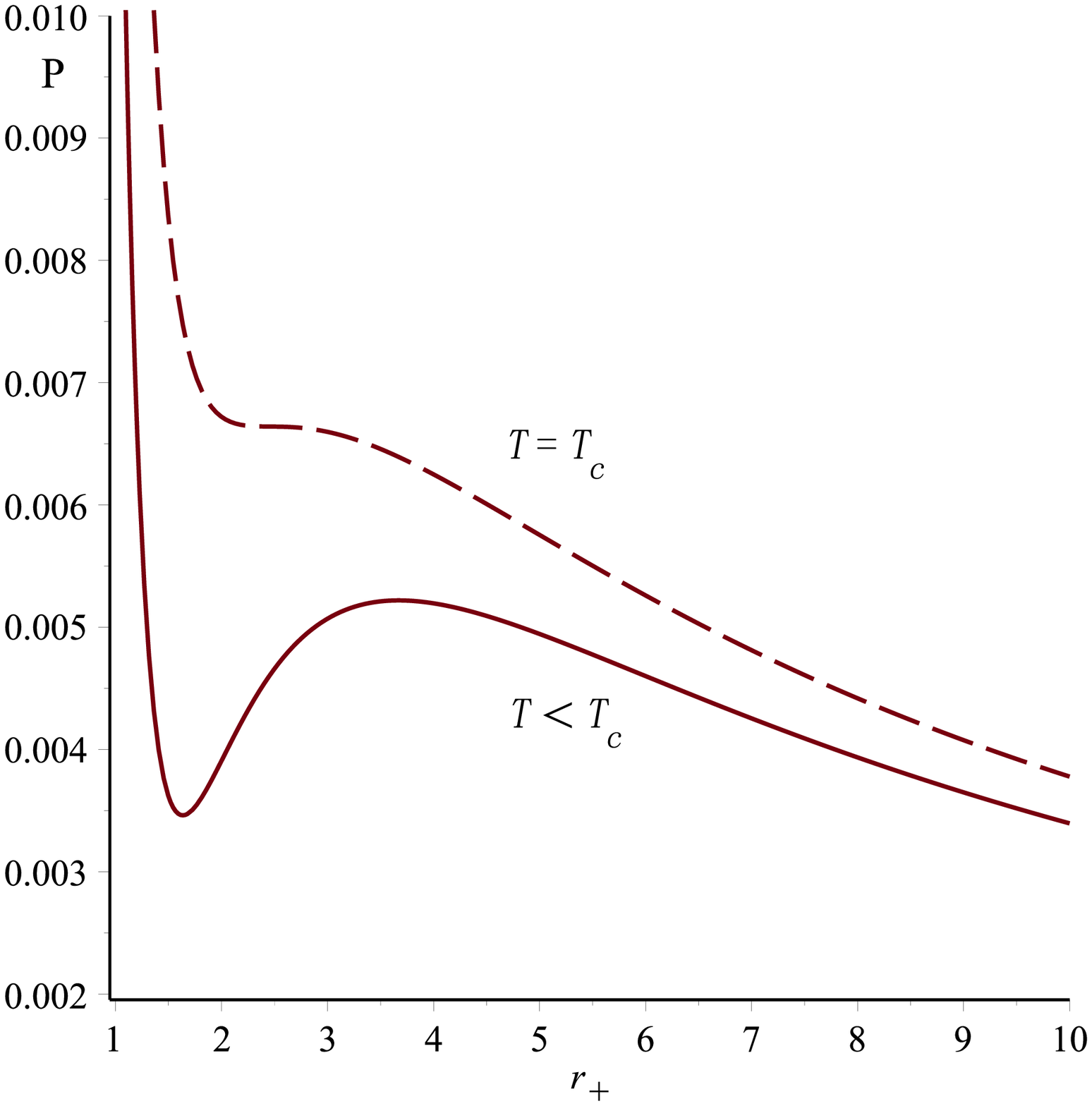}
\includegraphics[width=0.35\textwidth]{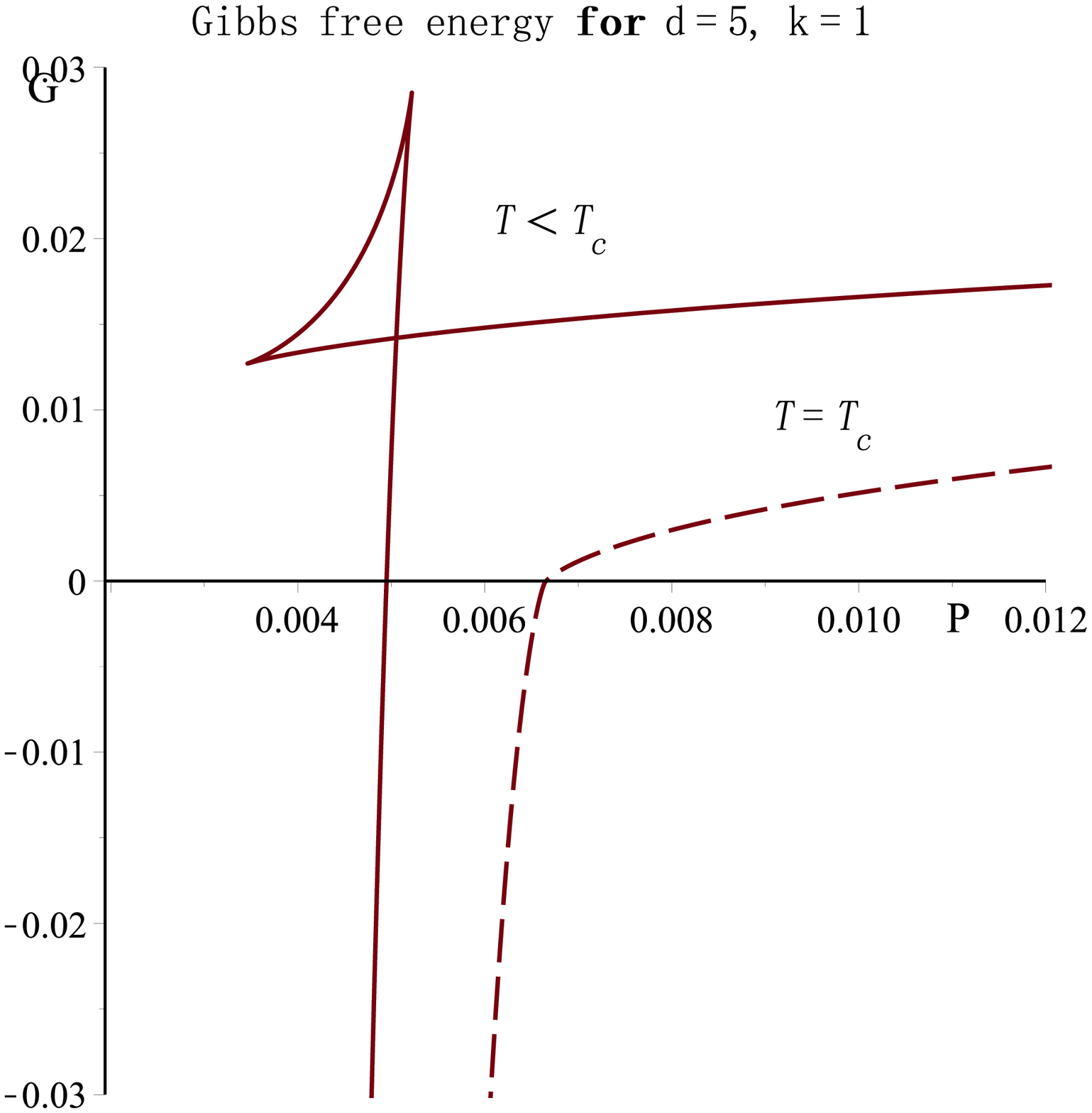}
\caption{Isothermal plots of the pressure and Gibbs free energy in $d=5$ and $k=+1$ for Gauss-Bonnet AdS black holes. The curves at $T=T_{c}$ are
depicted in dashed line. When $T<T_c$, there exist three black holes with different sizes at the same pressure, and the Gibbs free energy with respect to pressure develops a swallow tail which predicts a first order phase transition.}
\label{fig1}
\end{center}
\end{figure}

\begin{figure}
\begin{center}
\includegraphics[width=0.70\textwidth]{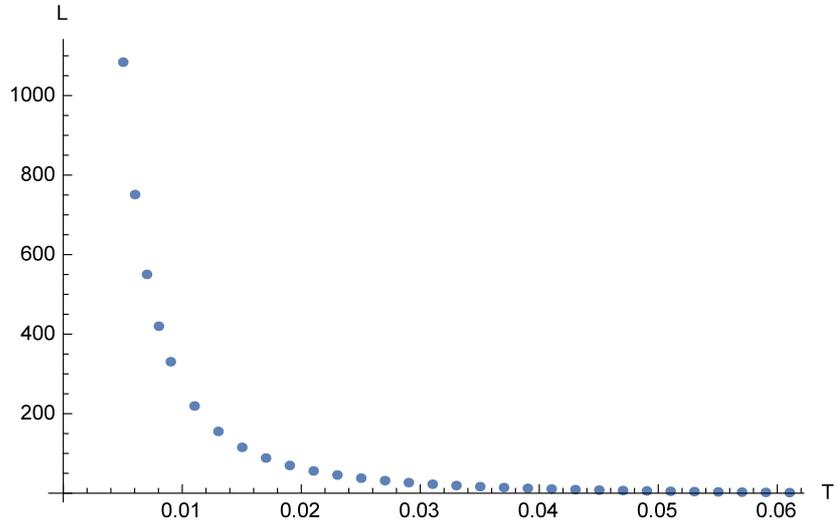}
\caption{Relationship between the latent heat $L$ and temperature $T$ in $d=5$ and $k=+1$ for Gauss-Bonnet AdS black holes. When $T\rightarrow 0$, $L$ is divergent.}
\label{fig2}
\end{center}
\end{figure}

If the dimension of the spacetime becomes higher, the analytic solution of \eqref{law} will become more difficult to obtain. In the following part we will use the numerical method to solve \eqref{law} by setting $\alpha=\Sigma_k=1$. If we fix the value of $T$, the radius $r_1$ and $r_2$ can be obtained directly. In Table \ref{tab1} and Figure \ref{fig1} we present some numerical solutions and isothermal plots of the pressure and Gibbs free energy. In Figure \ref{fig2} we give the plot of latent heat varies with temperature $T$. When $T$ is near to critical temperature $T_c$, the latent heat approaches zero. When $T$ decreases, the latent heat becomes larger. It goes to infinity as $T\rightarrow 0$. This also agrees well with our numerical analysis.

Secondly, when $d=6$, no real root of \eqref{law} can be found except $r_1=r_2=const$, so there is no critical behavior. This is because the last term in \eqref{P} dominates the equation of state at small radius. There do not exists the oscillating part where the phase transition could occur.

\begin{table}[!htbp]
\centering
\begin{tabular}{|c|c|c|c|c|}
\hline

$T$ ~&~  $r_1$ ~&~   $r_2$ ~&~ $G$  ~&~   $L$  \\
\hline
$0.065$ ~&~ $2.4495$ ~&~ $2.4495$ ~&~ $-0.0001$ ~&~ $0$  \\
\hline
$0.030$ ~&~ $0.4078$ ~&~ $14.7142$ ~&~ $0.0508$ ~&~ $24.5362$  \\
\hline
$0.010$ ~&~ $0.1267$ ~&~ $47.3671$ ~&~ $0.0587$ ~&~ $266.3965$  \\
\hline
$0.003$ ~&~ $0.0377$ ~&~ $159.0418$ ~&~ $0.0596$ ~&~ $3017.8521$  \\
\hline
$0.001$ ~&~ $0.0126$ ~&~ $477.4271$ ~&~ $0.0597$ ~&~ $27206.5026$  \\
\hline
\end{tabular}
\caption{The numerical solutions of Gauss-Bonnet black holes in $d=5$ and $k=+1$. We set $\alpha=\Sigma_k=1$. When $T=0.065=T_{c}$, we have $r_1=r_2=2.4495=r_c$. There is no first order phase transition. When $T<T_c$, we can obtain the value of $r_1$ and $r_2$. They also share the same Gibbs free energy and the latent heat can be obtained. When $T\rightarrow 0$, the latent heat is divergent.}
\label{tab1}
\end{table}

\section{Equal area law for the third order Lovelock black holes}
Now we study the equal area law for the third order Lovelock black holes. The action is given by
\begin{align}
{\cal I}=\frac{1}{16\pi}\int \mathrm{d}^{d}x\sqrt{-g}(R-2\Lambda+\alpha_{2}{\cal L}_{2}
+\alpha_{3}{\cal L}_{3}),
\end{align}
where the Gauss-Bonnet density ${\cal L}_2$ and the third order Lovelock density ${\cal L}_3$ read as
\begin{align}
{\cal L}_{2}=R_{\mu\nu\gamma\delta}R^{\mu\nu\gamma\delta}
-4R_{\mu\nu}R^{\mu\nu}+R^2,
\end{align}
\begin{align}
{\cal L}_{3}&=R^3+2R^{\mu\nu\sigma\kappa}R_{\sigma\kappa\rho\tau}
R^{\rho\tau}_{~~\mu\nu}
+8R^{\mu\nu}_{~~\sigma\rho}R^{\sigma\kappa}_{~~\nu\tau}
R^{\rho\tau}_{~~\mu\kappa}+24R^{\mu\nu\sigma\kappa}R_{\sigma\kappa\nu\rho}
R^{\rho}_{\mu}\nonumber\\
&\quad+3RR^{\mu\nu\sigma\kappa}R_{\mu\nu\sigma\kappa}
+24R^{\mu\nu\sigma\kappa}R_{\sigma\mu}R_{\kappa\nu}
+16R^{\mu\nu}R_{\nu\sigma}R^{\sigma}_{~\mu}-12RR^{\mu\nu}R_{\mu\nu}.
\end{align}

For the Gauss-Bonnet coefficients $\alpha_2$ and the third order Lovelock coefficients $\alpha_3$, we adapt the popular choice, which satisfy the following equations
\begin{align}
\alpha_{2}=\frac{\alpha}{(d-3)(d-4)},\quad
\alpha_{3}=\frac{\alpha^2}{72{d-3\choose 4}}.\label{eq:3a}
\end{align}
We can have the analytic static black hole solution of the form \cite{Dehghani:2005vh,Kofinas:2007ns,Dehghani:2009zzb,Zou:2010yr}
\begin{align}
ds^2=-f(r)\mathrm{d}t^2+\frac{1}{f(r)}\mathrm{d}r^2+r^2 \mathrm{d}
\Omega_k^2,\label{eq:2a}
\end{align}
\begin{align}
f(r)=k+\frac{r^2}{\alpha}\left(1-\left(1+\frac{6\Lambda \alpha}
{(d-1)(d-2)}+\frac{3\alpha m}
{r^{d-1}}\right)^\frac{1}{3}\right),
\end{align}

The radius of the black hole horizon is also the largest root of $f(r)=0$. We identify $H\equiv M$ and $T=\frac{f'(r_+)}{4\pi}$. Inserting $P=-\frac{\Lambda}{8\pi}$ and $M=\frac{(d-2)\Sigma_k m}{16\pi}$ into $f(r)$, we have
 \begin{align}
H&=\frac{(d-2)\Sigma_{k} r_+^{d-3}}{16\pi}
\left(k+\frac{16\pi P r_+^2}{(d-1)(d-2)}
+\frac{\alpha k^2}{r_+^2}+\frac{\alpha^2k}{3r_+^4}\right),\\
\label{eq.T}
T&=\frac{1}{12\pi r_+(r_+^2+k\alpha)^2}\bigg(\frac{48\pi r_+^6 P}{(d-2)}
+3(d-3)r_+^4k+3(d-5)r_+^2\alpha k^2+(d-7)\alpha^2k\bigg).
\end{align}

The $d=7$ case is qualitatively different from other choices, because the last term in (\ref{eq.T}) vanishes when $d=7$. The critical behavior in $d=7$ is also distinguished from the cases of higher dimensions.

The other thermodynamic quantities we need can also be easily calculated\cite{Xu:2014tja}. The black hole entropy
\begin{align}
\label{eq.S}
S=\frac{\Sigma_{k}r_+^{d-2}}{4}\left(1+\frac{2(d-2)k \alpha}{(d-4)r_+^2}
+\frac{(d-2)k^2\alpha^2}{(d-6)r_+^4}\right)
\end{align}
and the thermodynamic volume
\begin{align}
V=\left(\frac{\partial H}{\partial P}\right)_{S,\alpha}=\frac{r^{d-1}\Sigma_k}{d-1}.
\end{align}

These thermodynamic quantities satisfy the first law of black
hole thermodynamics in the extended phase space (\ref{firstlaw}).

Eq.\eqref{eq.T} can be rearranged into the following form,
\begin{align}
P&=\frac{T(d-2)}{4r_+}-\frac{k(d-2)(d-3)}{16\pi r_+^2}
+\frac{Tk\alpha (d-2)}{2r_+^3}-\frac{k^2 \alpha (d-2)(d-5)}{16\pi r+^4}
\nonumber\\
&\quad+\frac{Tk^2 \alpha^2 (d-2)}{4r_+^5}
-\frac{k^3\alpha^2 (d-2)(d-7)}{\pi r_+^6}. \label{eq.P}
\end{align}

Considering the Maxwell's equal area law described by (\ref{law}), we need to discuss following three cases.

\subsubsection*{1) Ricci flat case with $k=0$}
When $k=0$, \eqref{eq.P} reduces into
\begin{align}
P&=\frac{T(d-2)}{4r_+}
\end{align}
It is identical to the equation of state of an ideal gas, so there is no critical behavior
when $k=0$ in any dimension.

\subsubsection*{2) Hyperbolic case with $k=-1$}
In this case, the pressure will become
\begin{align}
P&=\frac{T(d-2)}{4r_+}+\frac{(d-2)(d-3)}{16\pi r_+^2}
-\frac{T\alpha (d-2)}{2r_+^3}-\frac{\alpha (d-2)(d-5)}{16\pi r+^4}
\nonumber\\
&\quad+\frac{T \alpha^2 (d-2)}{4r_+^5}
+\frac{\alpha^2 (d-2)(d-7)}{\pi r_+^6}.
\end{align}

\begin{figure}
\begin{center}
\includegraphics[width=0.38\textwidth]{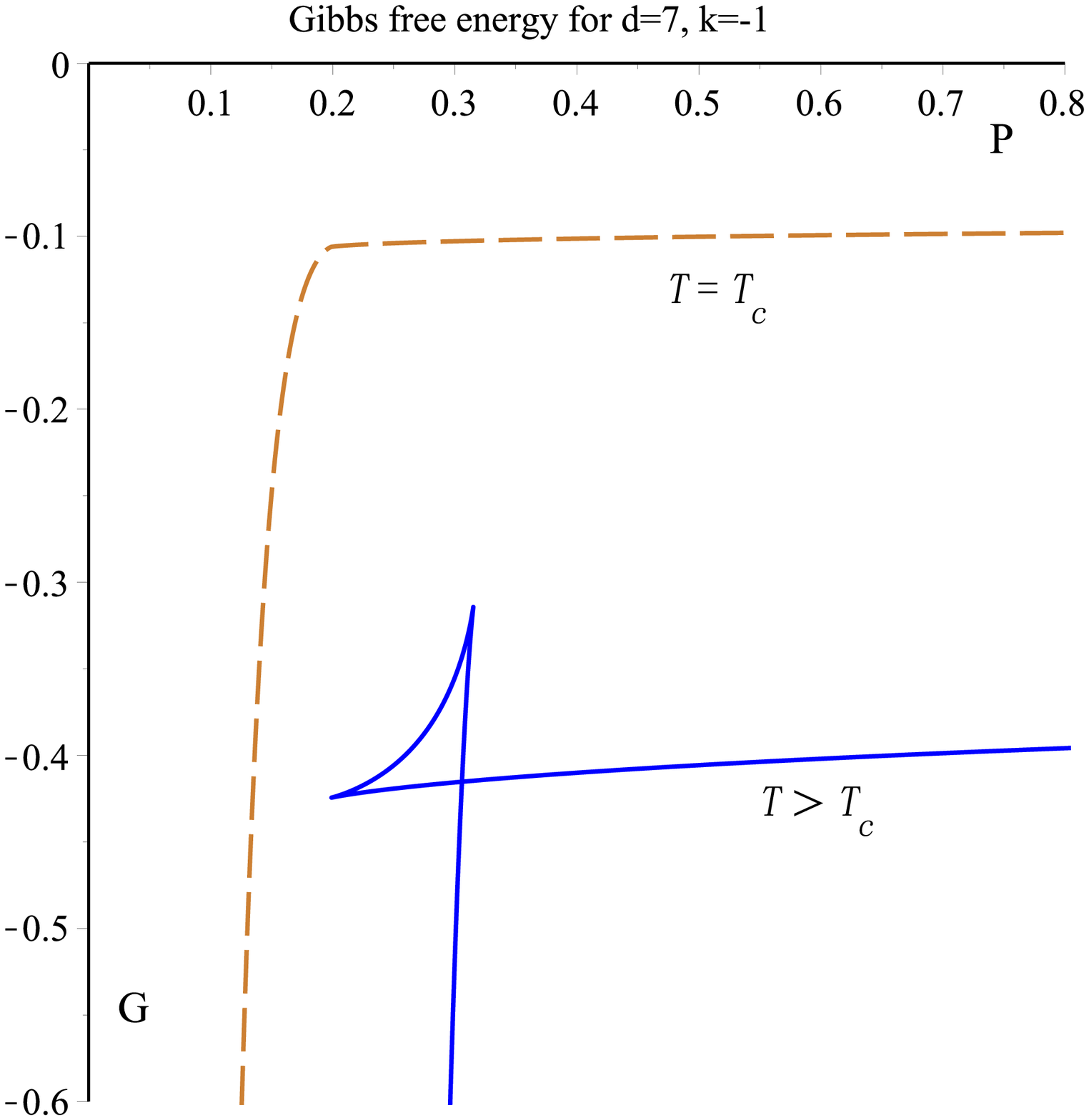}
\includegraphics[width=0.38\textwidth]{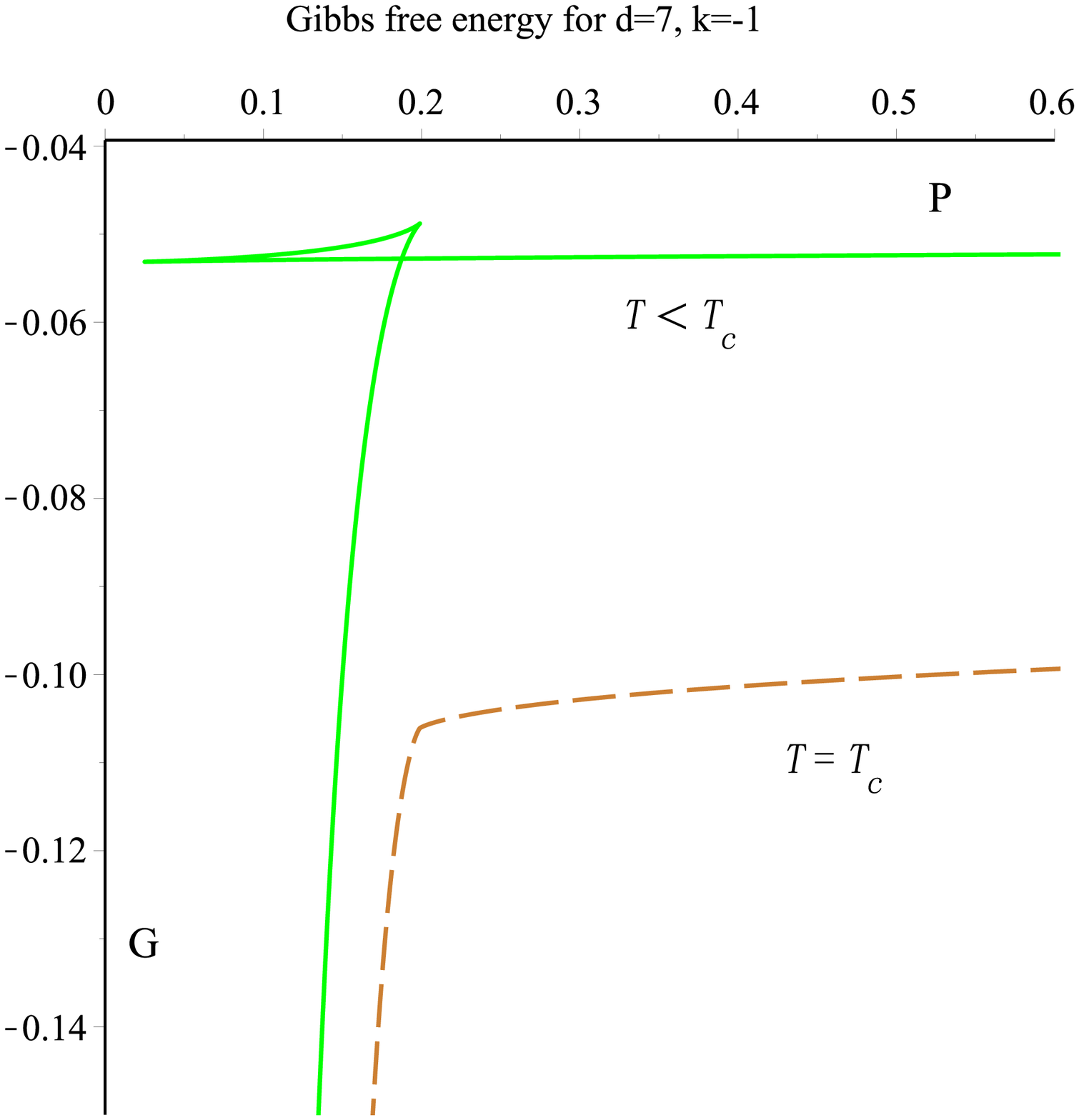}
\caption{Isothermal plots of Gibbs free energy at $d=7$ and $k=-1$ in $T>T_c$ and $T<T_c$. The curves at $T=T_{c}$ are
depicted in dashed line.}
\label{fig3}
\end{center}
\end{figure}

\begin{figure}
\begin{center}
\includegraphics[width=0.7\textwidth]{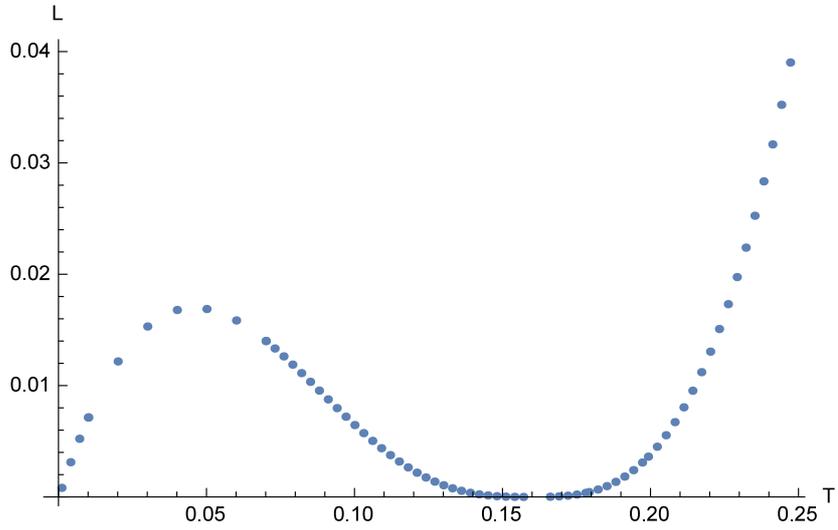}
\caption{Relationship between the latent heat $L$ and temperature $T$ in $d=7$ and $k=-1$ for third order Lovelock AdS black holes.}
\label{fig4}
\end{center}
\end{figure}

Firstly, let us consider the case of $d=7$ and $\alpha=1$. The last term in \eqref{P7} vanishes and the pressure takes the form
\begin{align}
P&=\frac{5T}{4r_+}+\frac{5}{4\pi r_+^2}
-\frac{5T}{2r_+^3}-\frac{(d-2)}{8\pi r+^4}+\frac{5T}{4r_+^5},
\label{P7}
\end{align}
and the thermodynamic volume becomes
\begin{align}
V=\frac{\Sigma_k r_+^{6}}{6}.
\label{V7}
\end{align}
Inserting the \eqref{P7} and \eqref{V7} into the equal area law (\ref{law}), we can solve the equations \eqref{law} by applying the numerical method. When $T=\frac{1}{2 \pi}$, we will have $r_1=r_2=1$. It is necessary to state $T=\frac{1}{2 \pi}$ and $r_+=1$ are exactly the critical temperature $T_c$ and critical radius $r_c$ obtained in \cite{Xu:2014tja} respectively. When $T<T_c$ or $T>T_c$, we can always find one solution $(r_1, r_2)$, which predicts the existence of a first order phase transition. This is very different from other cases, such as the RN AdS black hole, where the oscillating part in isothermal plot of pressure is only found in $T<T_c$, or four-dimensional conformal AdS black hole \cite{Xu:2014kwa}, where the phase transition happens in $T>T_c$. In Table \ref{tab2} and Figure \ref{fig3} we give the numerical solutions and isothermal plots of Gibbs free energy. In Figure \ref{fig4} we give the plot of latent heat varies with temperature $T$. The relationship between the latent heat and temperature is very different from other AdS black holes. The latent heat $L=0$ when $T=T_c$. In the region $T>T_c$, when $T$ increases, the latent heat also increases. It goes to infinity as $T$ approaches infinity. On the other hand, $L\rightarrow 0$ as $T\rightarrow 0$. There is an extreme value of $L$ in $0<T<T_c$. We do not know any other black holes which exhibit similar thermodynamic behaviors. It is would be interesting if we can find other examples which appear similar behaviors.

\begin{table}[!htbp]
\centering
\begin{tabular}{|c|c|c|c|c|}
\hline
$T$ ~&~  $r_1$ ~&~   $r_2$ ~&~ $G$  ~&~   $L$  \\
\hline
$10$ ~&~ $0.8247$ ~&~ $2.6743$ ~&~ $-6.4137$ ~&~ $209.4109$  \\
\hline
$5$ ~&~ $0.8262$ ~&~ $2.6327$ ~&~ $-3.2117$ ~&~ $95.2152$  \\
\hline
$1$ ~&~ $0.8398$ ~&~ $2.3184$ ~&~ $-0.6483$ ~&~ $8.5983$  \\
\hline
$\frac{1}{2 \pi}$ ~&~ $1$ ~&~ $1$ ~&~ $-\frac{1}{3 \pi}$ ~&~ $0$  \\
\hline
$0.1$ ~&~ $0.6339$ ~&~ $1.1249$ ~&~ $-0.0677$ ~&~ $0.0065$  \\
\hline
$0.05$ ~&~ $0.3146$ ~&~ $1.2682$ ~&~ $-0.0427$ ~&~ $0.0169$  \\
\hline
$0.01$ ~&~ $0.0628$ ~&~ $1.3857$ ~&~ $-0.0335$ ~&~ $0.0071$  \\
\hline
\end{tabular}
\caption{The numerical solutions of the third order Lovelock black holes in $d=7$ and $k=-1$. When $T=\frac{1}{2\pi}$, we have $r_1=r_2=1=r_c$. When $T\neq T_c$, there is always a swallow tail which predicts the first order phase transition.}
\label{tab2}
\end{table}

The results for the $d>7$ cases are extremely similar to the $d=7$ case. We can always find the critical temperature $T_c$ where $r_1$ and $r_2$ coincide. When $T\neq T_c$, there is a first order phase transition.

\subsubsection*{3) Spherical case with $k=+1$}
We also consider the case of $d=7$ first. In this case Eq.\eqref{eq.P} becomes
\begin{align}
P&=\frac{5T}{4r_+}-\frac{5}{4\pi r_+^2}
+\frac{5T}{2r_+^3}-\frac{5}{8\pi r+^4}+\frac{5T}{4r_+^5}.
\label{eq.P1}
\end{align}
 Inserting the pressure \eqref{eq.P1} and thermodynamic volume \eqref{V7} into the equal area law \eqref{law},
we find when $T>0.1424$, no real root can be found. $T=0.1424$ is the critical temperature $T_c$ obtained in \cite{Xu:2014tja}. In $T=0.1424$, we have $r_1=r_2=2.2361$, which is the critical radius $r_c$ \cite{Xu:2014tja}. When $T<0.1424$, we can find one solution $(r_1,r_2)$. There is a first order phase transition. The numerical solutions are given in Table \ref{tab3}. In Figure \ref{fig6} we give the relationship between the latent heat $L$ and temperature $T$. As $T$ is close to critical temperature $T_c$, the latent heat approaches zero. When $T$ decreases, the latent heat becomes larger. It goes to infinity as $T\rightarrow 0$. This is very similar with the Gauss Bonnet AdS black holes in $d=5$ and $k=+1$.

\begin{table}[!htbp]
\centering
\begin{tabular}{|c|c|c|c|c|}
\hline

$T$ ~&~  $r_1$ ~&~   $r_2$ ~&~
$G$  ~&~   $L$  \\
\hline
$0.1424$ ~&~ $2.2361$ ~&~ $2.2361$ ~&~ $-0.1331$ ~&~ $0$  \\
\hline
$0.1000$ ~&~ $0.7000$ ~&~ $6.7497$ ~&~ $-0.0143$ ~&~ $376.5958$  \\
\hline
$0.0500$ ~&~ $0.3168$ ~&~ $15.3934$ ~&~ $0.0230$ ~&~ $1.0956\times 10^4$  \\
\hline
$0.0100$ ~&~ $0.0628$ ~&~ $79.4768$ ~&~ $0.0328$ ~&~ $7.9318\times 10^6$  \\
\hline
\end{tabular}
\caption{The numerical solutions of the third order Lovelock black holes in $d=7$ and $k=+1$. When $T<0.1424$, there is a first order phase transition. The latent heat $L\rightarrow +\infty$ when $T\rightarrow 0$.}
\label{tab3}
\end{table}

\begin{figure}
\begin{center}
\includegraphics[width=0.35\textwidth]{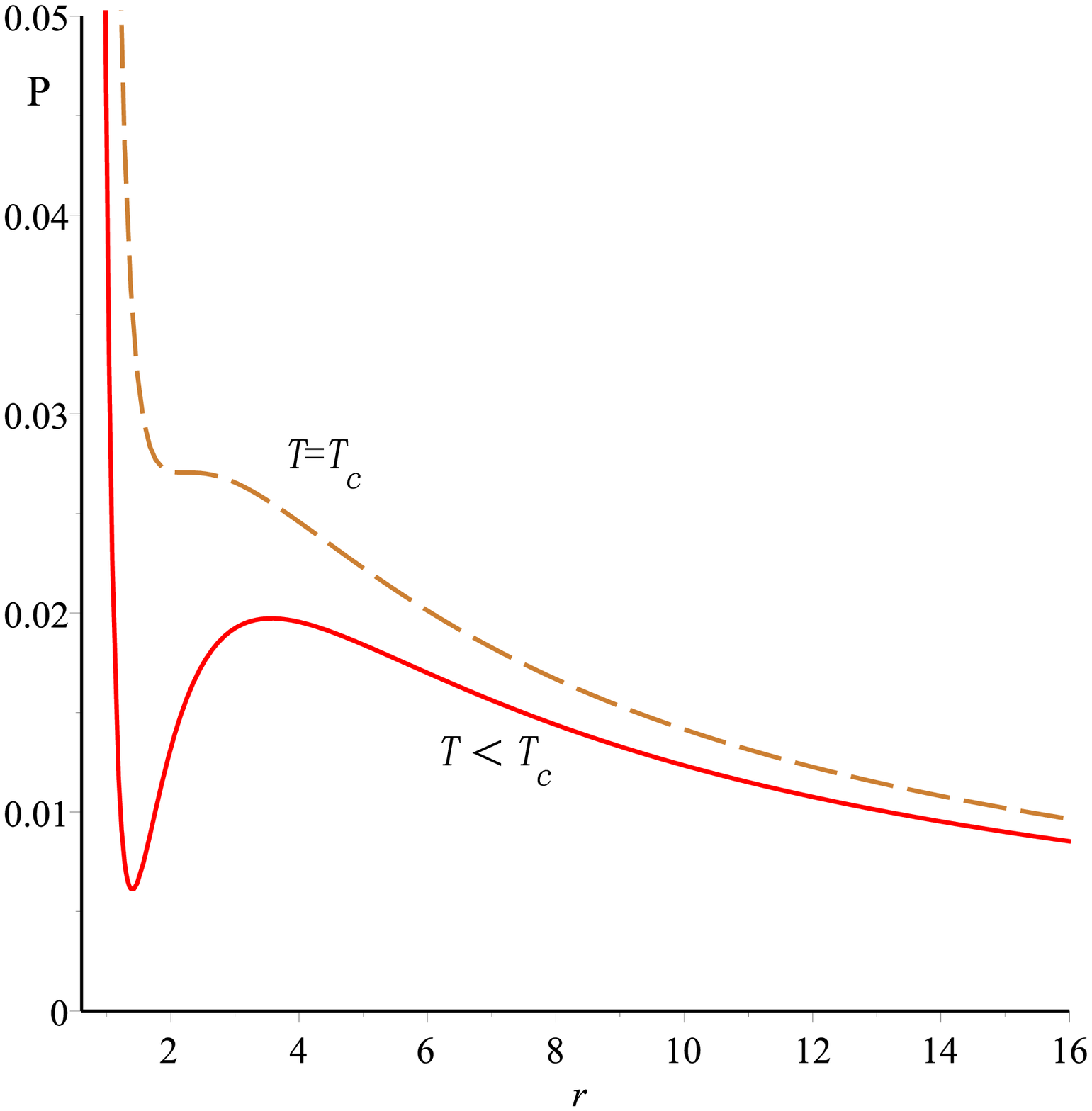}
\includegraphics[width=0.35\textwidth]{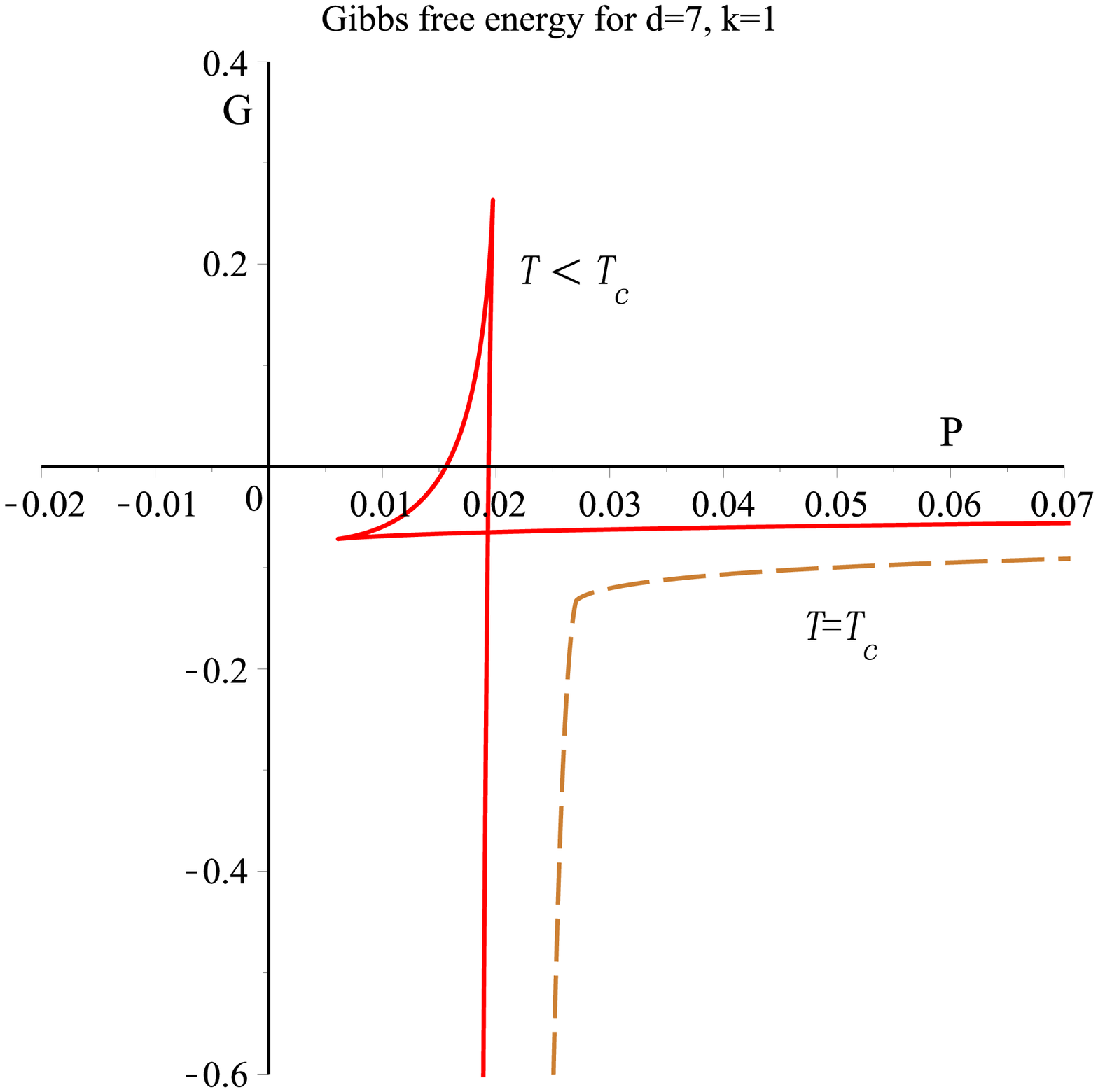}
\caption{Isothermal plots of the pressure and Gibbs free energy at $d=7$ and $k=+1$ for third order Lovelock black holes. The curves at $T=T_{c}$ are
depicted in dashed line.}
\label{fig5}
\end{center}
\end{figure}

\begin{figure}
\begin{center}
\includegraphics[width=0.7\textwidth]{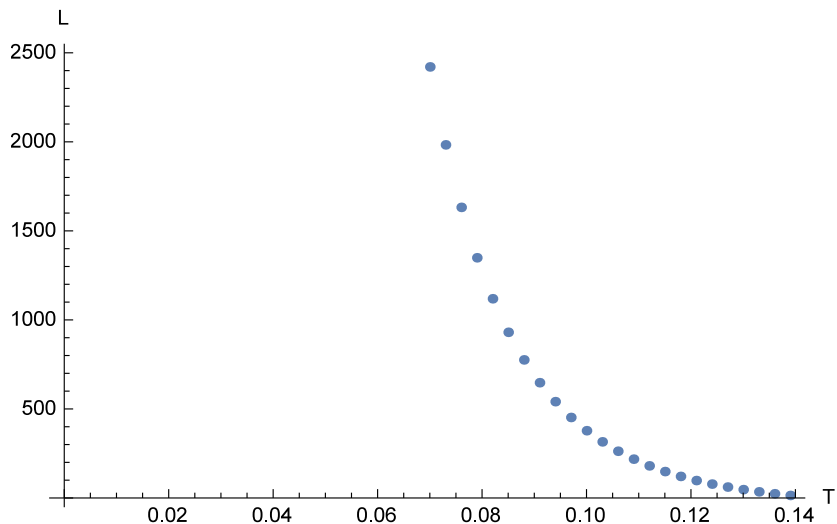}
\caption{Relationship between the latent heat $L$ and temperature $T$ in $d=7$ and $k=+1$ in third order Lovelock AdS black holes.}
\label{fig6}
\end{center}
\end{figure}

Next we investigate the case $d=8$, which is much more complicated. Numerical solutions show that when $T<0.1364$ or $T>0.1857$, no real root can be found in the equation of
equal area law. $T=0.1364$ and $T=0.1857$ are the two critical points labeled $T_{c1}$ and $T_{c2}$ found in \cite{Xu:2014tja} respectively. The Gibbs free energy
resembles the curve characteristic for the Schwarzschild AdS black hole.

In $T=T_c$, the isobar which represents the coexistence of small and large black holes shrinks to a point. When temperature becomes larger or(and)
smaller than the $T_c$, the oscillating part in the isothermal plot of pressure begins to appear. However, in our system, there are two critical temperature and the swallow tail can be found when temperature is near both of them. In the region $T_{c1}<T<T_{c2}$, we have one solution as $T$ approaches $T_{c1}$ or $T_{c2}$ and three solutions in the middle of it. The isothermal plots of Gibbs free energy are complicated. By solving the equations of equal area law, we can obtain the number and location of the points of intersection in the isothermal plots of the Gibbs free energy, so that we can know where the phase transition appears and the value of latent heat.

\begin{table}[!htbp]
\centering
\begin{tabular}{|c|c|c|c|c|}
\hline
$T$ ~&~  $r_1$ ~&~   $r_2$ ~&~  $G$  ~&~   $L$  \\
\hline
$0.1390$ ~&~ $0.3540$ ~&~ $0.7085$ ~&~ $0.0052$ ~&~ none  \\
\hline
$0.1393$ ~&~ $0.3493$ ~&~ $0.7204$ ~&~ $0.0052$ ~&~ none  \\
$ $ ~&~ $0.3924$ ~&~ $5.6093$ ~&~ $0.0053$ ~&~ $1191.4823$ \\
\hline
$0.1580$ ~&~ $0.2297$ ~&~ $1.2802$ ~&~ $0.0041$ ~&~ none  \\
$$ ~&~ $0.2298$ ~&~ $4.4869$ ~&~ $0.0040$ ~&~ none  \\
$ $ ~&~ $0.7440$ ~&~ $4.4877$ ~&~ $-0.0020$ ~&~ $372.9974$  \\
\hline
$0.1825$ ~&~ $1.4309$ ~&~ $2.7940$ ~&~ $-0.0488$ ~&~ $29.8691$  \\
\hline
\end{tabular}
\caption{The numerical solutions of the third order Lovelock black holes in $d=8$. 'None' means that there is a solution which predicts a point of intersection in the isothermal plots of Gibbs free energy, but it is not thermodynamically preferred, which does not predict a first order phase transition.}
\label{tab4}
\end{table}

\begin{figure}[h!!]
\begin{center}
\includegraphics[width=0.35\textwidth]{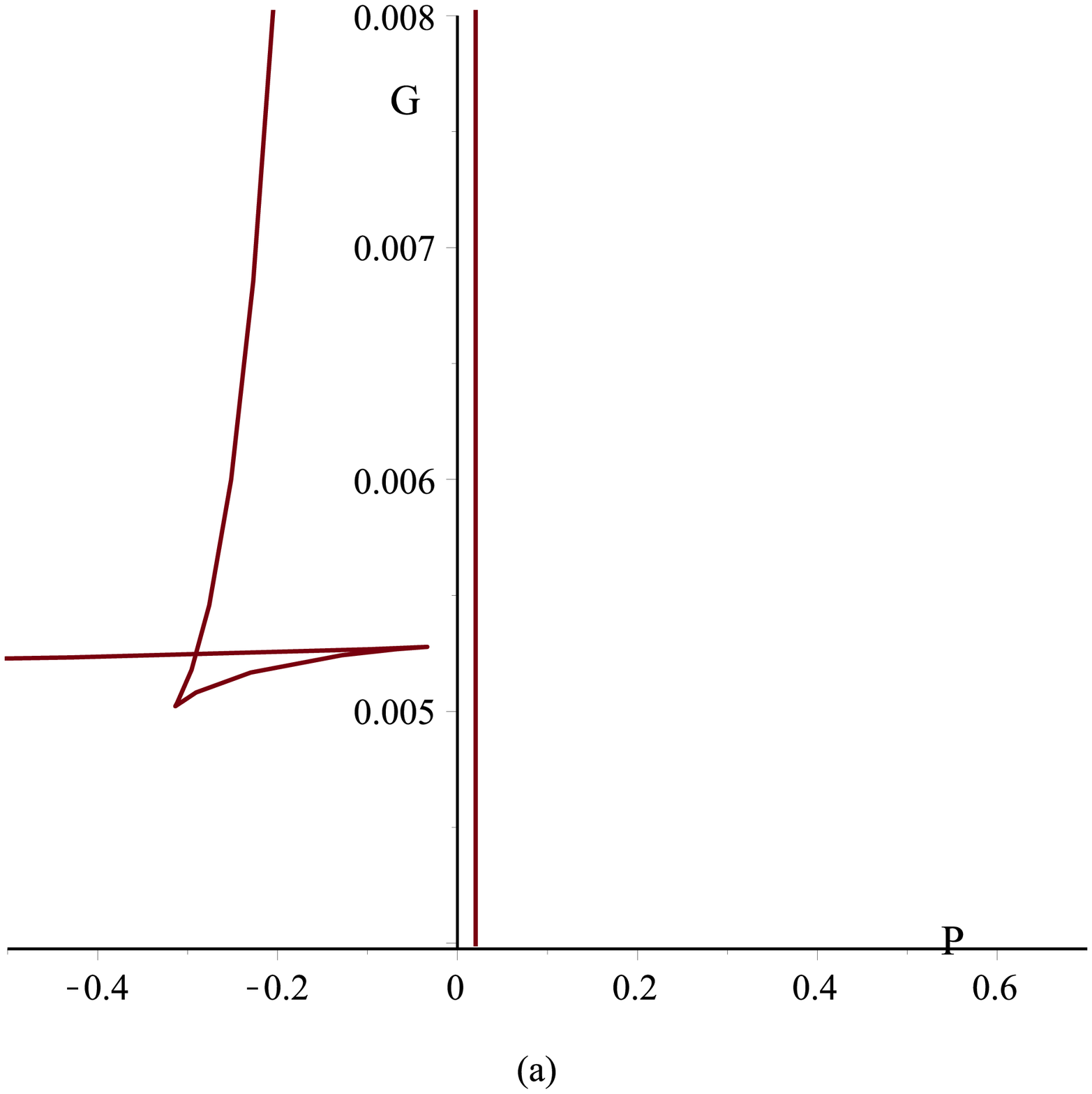}
\includegraphics[width=0.35\textwidth]{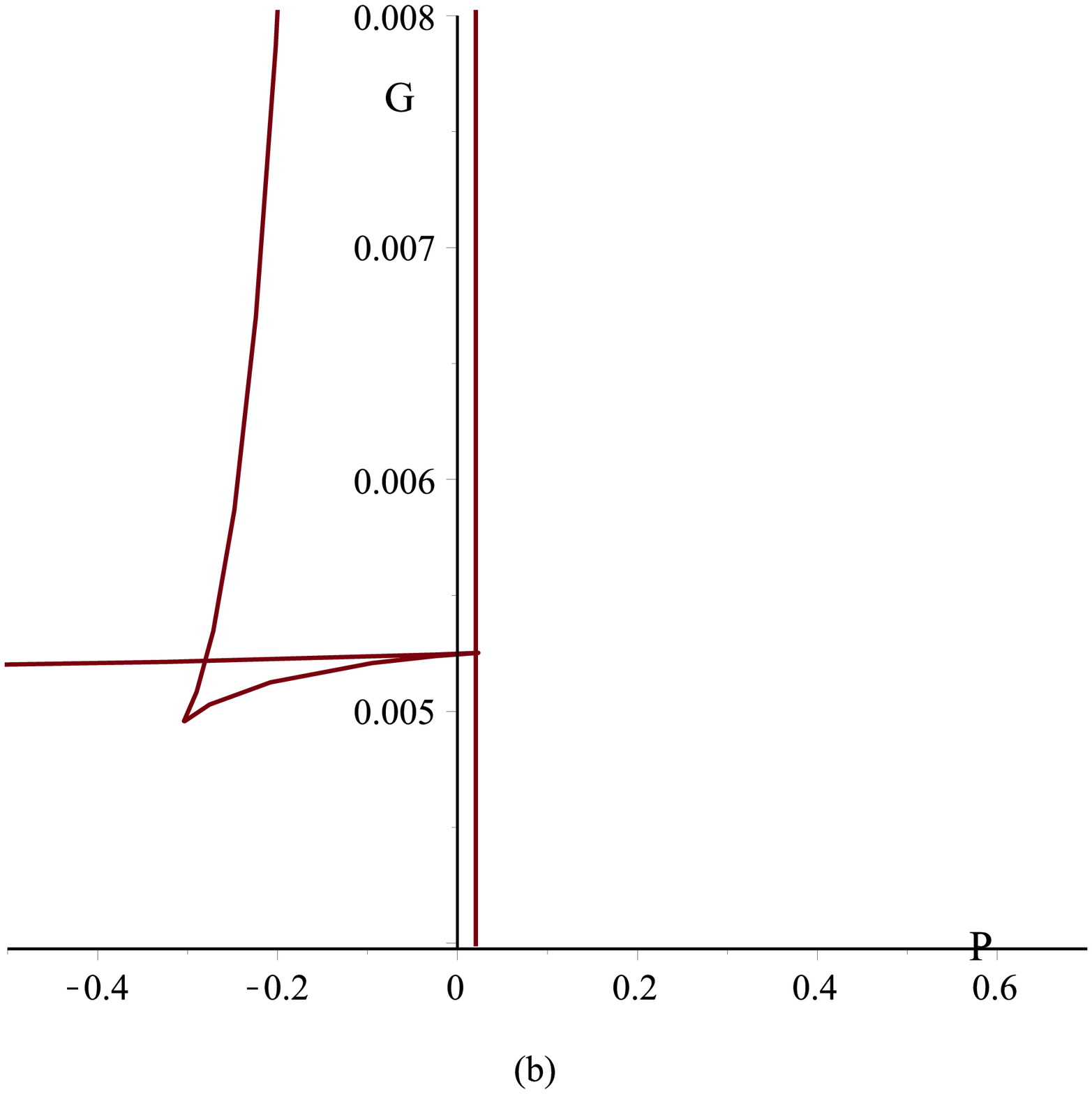}
\includegraphics[width=0.35\textwidth]{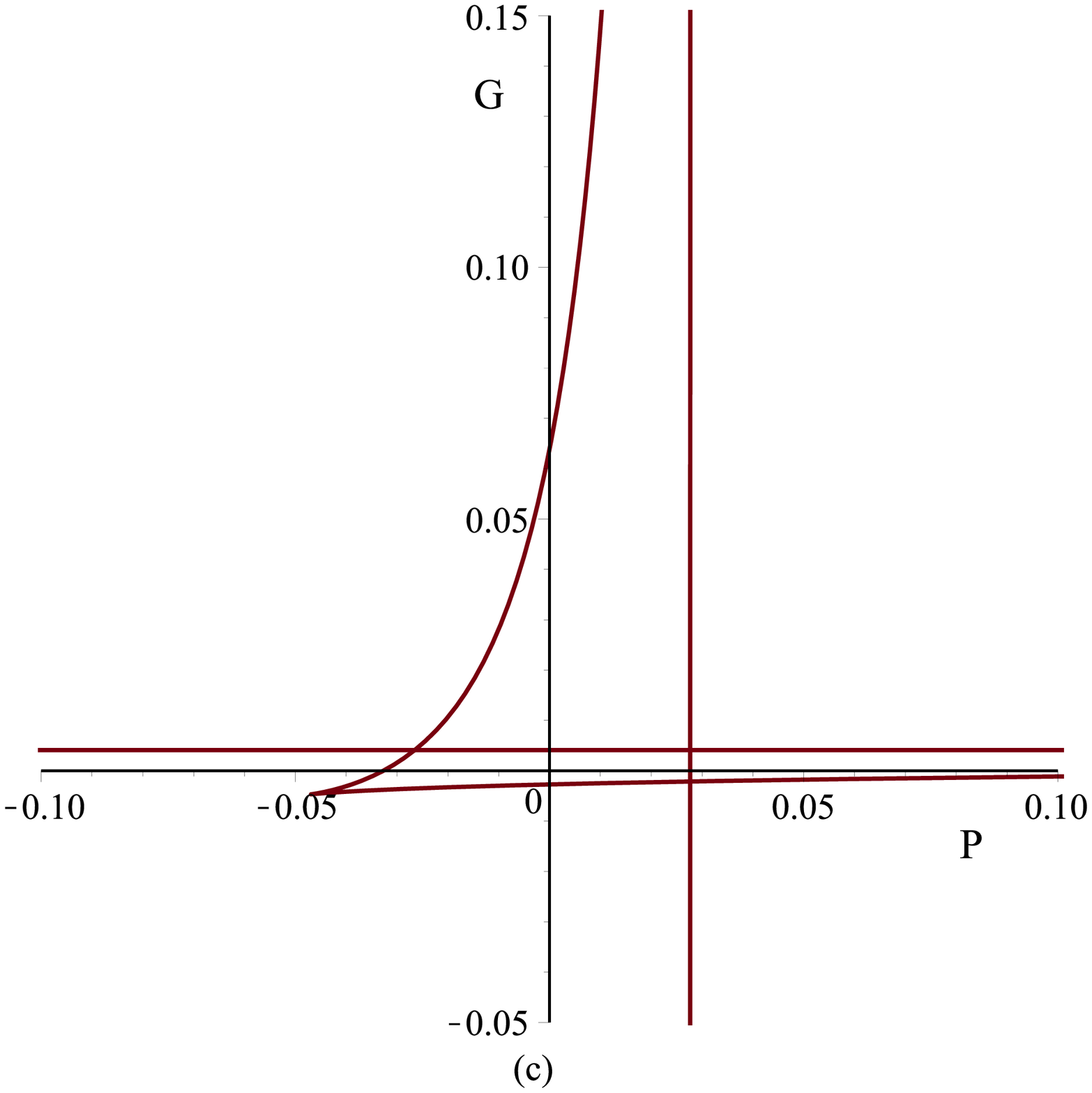}
\includegraphics[width=0.35\textwidth]{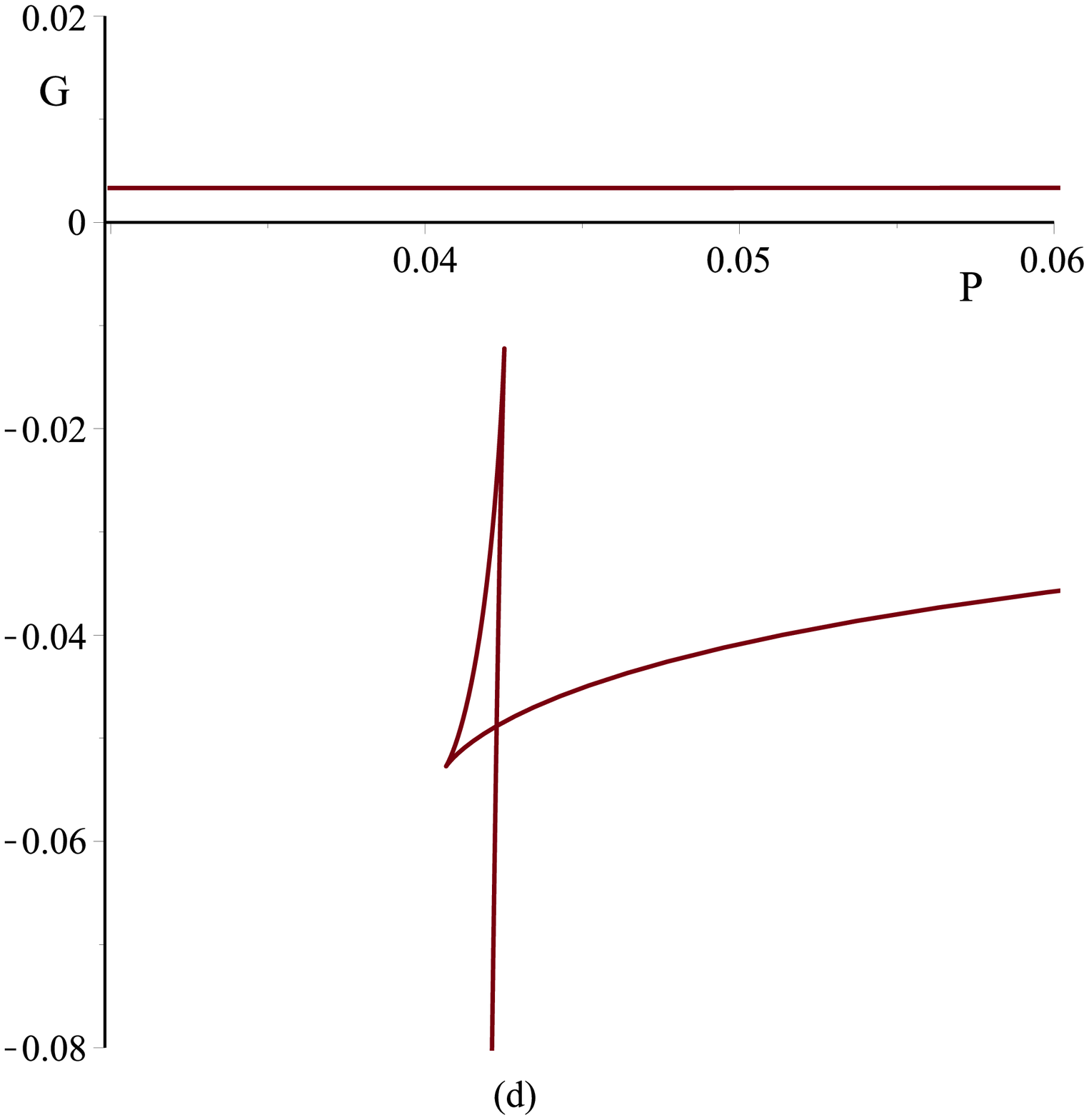}
\caption{Isothermal plots of the Gibbs free energy at $d=8$ and $k=+1$
at $T_{c1}<T<T_{c2}$. These figures correspond to $T=0.1390$, $T=0.1393$, $T=0.1580$ and $T=0.1825$ respectively. We pay our attention to the region where the solutions of equal area law are found. In (a), (b) and (c) the lines along the $G$ direction would join together in a large but limited $G$. The lines in (c) and (d) along the $P$ direction also join together in a large but limited $P$. As $r_+\rightarrow \infty$, $P\rightarrow 0$.}
\label{fig7}
\end{center}
\end{figure}

\begin{figure}
\begin{center}
\includegraphics[width=0.7\textwidth]{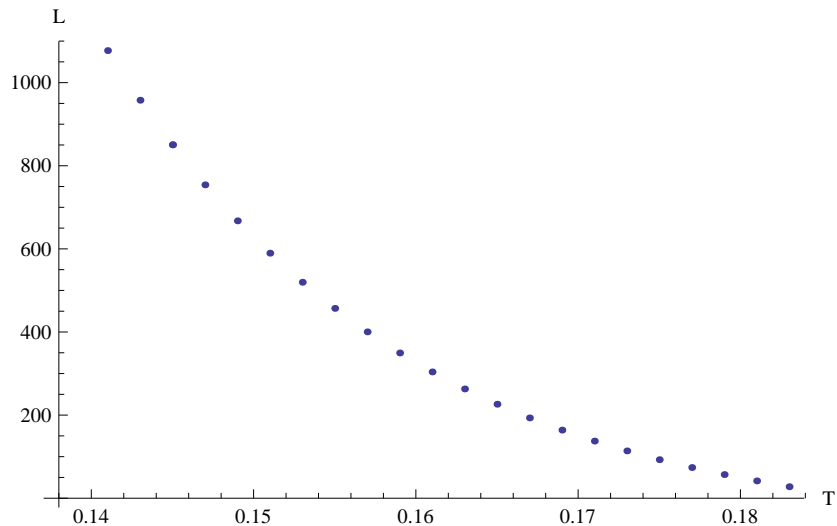}
\caption{Relationship between the latent heat $L$ and temperature $T$ in $d=8$ and $k=1$ in third order Lovelock AdS black holes.}
\label{fig8}
\end{center}
\end{figure}

In Table \ref{tab4} and Figure \ref{fig4} we give the numerical solutions  and corresponding isothermal plots of Gibbs free energy. When $T=0.1390$, which is slightly larger than $T=T_{c1}$, we have one solution of $(r_1,r_2)$. However, we do not consider it as a phase transition for two reasons. Firstly, the corresponding pressure $P<0$. The pressure $P=-\frac{\Lambda}{8 \pi}$ can become negative in a isothermal process. However, since the $P-V$ analysis may not valid in de Sitter spacetime, we would consider the $P<0$ as unphysical in our paper. Secondly, when the temperature becomes higher, this solution can moved to the region $P>0$, but it is only locally stable and not thermodynamically preferred. This is similar with the Hawking-Page phase transition. When $T=0.1393$, the swallow tail corresponding to the $T_{c2}$ and the first order phase transition begin to appear. We denote the $T=0.1393$ as $T_t$. In fact, when the temperature is slightly larger than $T_t$, we can observe a finite jump in Gibbs free energy, which has a discontinuous global minimum. This is the reentrant phase transition, a phenomenon seen in rotating black holes\cite{Altamirano:2013ane} and Born-Infeld black holes \cite{Gunasekaran:2012dq}. At $T=0.1580$ we can have three solutions because the two swallow tails cross each other. Two of the solutions are not thermodynamically preferred while the other one predicts the phase transition. When $T=0.1825$, the swallow tail corresponding to $T_{c1}$ has already disappeared. We only have one solution, which also predict the first order phase transition. In the region $T_t<T<T_{c2}$, rather than $T_{c1}<T<T_{c2}$, we can find the first order phase transition, so the $T_t$ is the 'real critical temperature' which can predict the thermodynamically preferred first order phase transition, while the the smaller critical point $T_{c1}$ does not participate in the phase transition. In Figure \ref{fig8} we give the relationship between the latent heat $L$ and temperature $T$ in the region $T_t<T<T_{c2}$. When $T$ increases, the latent heat decreases. The latent heat $L$ vanishes as $T=T_{c2}$. When $T<0.1393$, there is no latent heat.

In the cases of $d=9,10,11$, the calculation will be much more tedious. All the solutions of $r_1$ and $r_2$ lie in the region of $P>0$ in $d=10,11$. This is the only difference with $d=8,9$. Other results are similar with the case of $d=8$. We will not present it here. When $d\geq 12$, there is no critical behavior.

\section{Concluding remarks}

Although the Maxwell's equal area law for some systems have been studied, other systems with rich structures are somehow overlooked. In this paper, we explored the equal area law for the Gauss Bonnet and third order Lovelock gravity, there are several interesting features:

\begin{itemize}
\item In the $5-d$ Gauss Bonnet AdS black holes and $7-d$ third order Lovelock black holes at $k=+1$, there is one solution of Maxwell's equal area law when $T<T_{c}$, which predicts a first order phase transition. When $T\rightarrow 0$, the latent heat $L\rightarrow +\infty$;

\item In the third order Lovelock black holes when $k=-1$, there is always a solution when $T<T_c$ or $T>T_c$. when $T\rightarrow +\infty$, the latent heat $L$ is divergent. When $T\rightarrow 0$, the latent heat also approaches zero. There is an extreme value of $L$ in $0<T<T_c$. We do not know any other black holes which exhibit similar thermodynamic behaviors;

\item When there are two critical points, such as the third order Lovelock black hole in $d=8$ and $k=+1$, the reentrant phase transition may appear. The number of the solutions of equal area law reveals the number of the points of intersection in the plots of Gibbs free energy. The first order phase transition can only be found in $T_t<T<T_{c1}$, where $T_t$ is the 'real critical temperature ' which predicts the thermodynamically preferred phase transition. The swallow tail corresponding to the $T_{c1}$ is only locally stable. One can also investigate the second order phase transition in these critical points\cite{Banerjee:2010da,Lala:2011np,Zhao:2014sra,Mo:2013ela,Mo:2014lza,Mo:2014wca,Mo:2014mba}, but they are not always globally stable. The latent heat decreases from $T_t$ to $T_{c2}$.
\end{itemize}

The Maxwell's equal area law provides us an efficient way to study the critical behavior of black holes. We can obtain the physical quantities of the black holes during the phase transition, such as the radius, Gibbs free energy and latent heat. It would be interesting to explore the use of equal area law in other gravity theories, especially the ones with unusual thermodynamical behaviors.

\section{Acknowledgment}
We thank Yuan Sun and Professor Liu Zhao for many related discussions.


\begin{thebibliography}{100}

\bibitem{HawkingPage:1983}
S.~Hawking and D.~N. Page, {\it {Thermodynamics of Black Holes in anti-De
  Sitter Space}},  {\em Commun.Math.Phys.} {\bf 87} (1983) 577.

\bibitem{Witten:1998zw}
  E.~Witten,
  Adv.\ Theor.\ Math.\ Phys.\  {\bf 2}, 505 (1998)
  [\eprint{hep-th/9803131}].

\bibitem{Witten:1998qj}
  E.~Witten,
  Adv.\ Theor.\ Math.\ Phys.\  {\bf 2}, 253 (1998)
   [\eprint{hep-th/9802150}].



\bibitem{ChamblinEtal:1999a}
A.~Chamblin, R.~Emparan, C.~Johnson, and R.~Myers,
   {\em Phys.Rev.} {\bf D60} (1999)
  064018, [\eprint{hep-th/9902170}].

\bibitem{ChamblinEtal:1999b}
A.~Chamblin, R.~Emparan, C.~V.~Johnson and R.~C.~Myers,
  Phys.\ Rev.\ D {\bf 60}, 104026 (1999)
  [\eprint{hep-th/9904197}].


\bibitem{Liu:2014gvf}
  Y.~Liu, D.~-C.~Zou and B.~Wang,
   JHEP {\bf 1409}, 179 (2014)
  [\eprint{hep-th/1405.2644}].


\bibitem{KastorEtal:2009}
 D.~Kastor, S.~Ray and J.~Traschen,
  Class.\ Quant.\ Grav.\  {\bf 26}, 195011 (2009)
  [\eprint{0904.2765}].


\bibitem{Dolan:2010}
  B.~P.~Dolan,
  Class.\ Quant.\ Grav.\  {\bf 28}, 125020 (2011)
  [\eprint{1008.5023}].



\bibitem{Dolan:2011a}
  B.~P.~Dolan,
  Class.\ Quant.\ Grav.\  {\bf 28}, 235017 (2011)
  [\eprint{1106.6260}].


\bibitem{Dolan:2011b}
 B.~P.~Dolan,
  Phys.\ Rev.\ D {\bf 84}, 127503 (2011)
  [\eprint{1109.0198}].



\bibitem{CveticEtal:2011}
  M.~Cvetic, G.~W.~Gibbons, D.~Kubiznak and C.~N.~Pope,
  Phys.\ Rev.\ D {\bf 84}, 024037 (2011)
  [\eprint{1012.2888}].


\bibitem{Hennigar:2014cfa}
  R.~A.~Hennigar, D.~Kubiz¨¾¨¢k and R.~B.~Mann,
  Phys.\ Rev.\ Lett.\  {\bf 115}, no. 3, 031101 (2015)
  [\eprint{1411.4309}].

\bibitem{Brenna:2015pqa}
  W.~G.~Brenna, R.~B.~Mann and M.~Park,
  Phys.\ Rev.\ D {\bf 92}, no. 4, 044015 (2015)
    [\eprint{1505.06331}].

\bibitem{D.Kubiznak}
 D.~Kubiznak and R.~B.~Mann,
  JHEP {\bf 1207}, 033 (2012)
  [\eprint{1205.0559}].




\bibitem{Poshteh:2013pba}
  M.~B.~J.~Poshteh, B.~Mirza and Z.~Sherkatghanad,
  Phys.\  Rev.\ D {\bf 88}, 024005 (2013)
  [\eprint{1306.4516}].


\bibitem{Belhaj:2013cva}
  A.~Belhaj, M.~Chabab, H.~E.~Moumni, L.~Medari and M.~B.~Sedra,
  Chin.\ Phys.\ Lett.\  {\bf 30}, 090402 (2013)
  [\eprint{1307.7421}].


\bibitem{Altamirano:2013uqa}
  N.~Altamirano, D.~Kubiznak, R.~B.~Mann and Z.~Sherkatghanad,
  [\eprint{1308.2672}].


\bibitem{Altamirano:2013ane}
  N.~Altamirano, D.~Kubiznak and R.~B.~Mann,
  Phys.\ Rev.\ D {\bf 88}, 101502 (2013)
  [\eprint{1306.5756}].


\bibitem{Altamirano:2014tva}
  N.~Altamirano, D.~Kubiznak, R.~B.~Mann and Z.~Sherkatghanad,
   Class.\ Quant.\ Grav.\  {\bf 31} (2014) 042001
  [\eprint{1401.2586}].




\bibitem{Wei:2012ui}
 S.~-W.~Wei and Y.~-X.~Liu,
  Phys.\ Rev.\ D {\bf 87}, no. 4, 044014 (2013)
  [\eprint{1209.1707}].




\bibitem{Cai:2013qga}
  R.~-G.~Cai, L.~-M.~Cao, L.~Li and R.~-Q.~Yang,
  JHEP {\bf 1309}, 005 (2013)
  [\eprint{1306.6233}].


\bibitem{Zou:2014mha}
  D.~-C.~Zou, Y.~Liu and B.~Wang,
    Phys.\ Rev.\ D {\bf 90}, no. 4, 044063 (2014)
   [\eprint{1404.5194}].


\bibitem{Chen:2013ce}
  S.~Chen, X.~Liu, C.~Liu and J.~Jing,
  Chin.\  Phys.\  Lett.\  {\bf 30}, 060401 (2013)
  [\eprint{1301.3234}].



\bibitem{Hristov:2013sya}
  K.~Hristov, C.~Toldo and S.~Vandoren,
  Phys.\ Rev.\ D {\bf 88}, 026019 (2013)
  [\eprint{1304.5187}].



\bibitem{Belhaj:2013ioa}
  A.~Belhaj, M.~Chabab, H.~El Moumni and M.~B.~Sedra,
    Int.\ J.\ Geom.\ Meth.\ Mod.\ Phys.\  {\bf 12}, no. 02, 1550017 (2014)
  [\eprint{1306.2518}].



\bibitem{Hendi:2012um}
 S.~H.~Hendi and M.~H.~Vahidinia,
  Phys.\ Rev.\ D {\bf 88}, 084045 (2013)
  [\eprint{1212.6128}].




\bibitem{Gunasekaran:2012dq}
  S.~Gunasekaran, R.~B.~Mann and D.~Kubiznak,
  JHEP {\bf 1211}, 110 (2012)
  [\eprint{1208.6251}].




\bibitem{Zou:2013owa}
  D.~-C.~Zou, S.~-J.~Zhang and B.~Wang,
  Phys.\ Rev.\ D {\bf 89}, 044002 (2014)
  [\eprint{1311.7299}].


\bibitem{Ma:2013aqa}
  M.~-S.~Ma, H.~-H.~Zhao, L.~-C.~Zhang and R.~Zhao,
    Int.\ J.\ Mod.\ Phys.\ A {\bf 29}, 1450050 (2014)
  [\eprint{1312.0731}].




\bibitem{Mo:2014qsa}
  J.~-X.~Mo and W.~-B.~Liu,
  Eur.\ Phys.\ J.\ C {\bf 74} (2014) 2836
  [\eprint{1401.0785}].

\bibitem{Johnson:2014yja}
  C.~V.~Johnson,
  Class.\ Quant.\ Grav.\  {\bf 31}, 205002 (2014)
     [\eprint{1404.5982}].

\bibitem{Johnson:2014xza}
  C.~V.~Johnson,
  Class.\ Quant.\ Grav.\  {\bf 31}, no. 23, 235003 (2014)
  [\eprint{1405.5941}].

\bibitem{Johnson:2014pwa}
  C.~V.~Johnson,
  Class.\ Quant.\ Grav.\  {\bf 31}, 225005 (2014)
  [\eprint{1406.4533}].


\bibitem{Xu:2014kwa}
  W.~Xu and L.~Zhao,
  Phys.\ Lett.\ B {\bf 736}, 214 (2014)
    [\eprint{1405.7665}].

\bibitem{Frassino:2014pha}
  A.~M.~Frassino, D.~Kubiznak, R.~B.~Mann and F.~Simovic,
  JHEP {\bf 1409}, 080 (2014)
  [\eprint{1406.7015}].

\bibitem{Dolan:2014vba}
  B.~P.~Dolan, A.~Kostouki, D.~Kubiznak and R.~B.~Mann,
  Class.\ Quant.\ Grav.\  {\bf 31}, no. 24, 242001 (2014)
    [\eprint{1407.4783}].

\bibitem{Lee:2014tma}
  C.~O.~Lee,
  Phys.\ Lett.\ B {\bf 738}, 294 (2014)
  [\eprint{1408.2073}].

\bibitem{Zhang:2014uoa}
  J.~L.~Zhang, R.~G.~Cai and H.~Yu,
  JHEP {\bf 1502}, 143 (2015)
    [\eprint{1409.5305}].

\bibitem{Wei:2014qwa}
  S.~W.~Wei and Y.~X.~Liu,
  Phys.\ Rev.\ D {\bf 91}, no. 4, 044018 (2015)
      [\eprint{1411.5749}].

\bibitem{Zhang:2015ova}
  J.~L.~Zhang, R.~G.~Cai and H.~Yu,
  Phys.\ Rev.\ D {\bf 91}, no. 4, 044028 (2015)
  [\eprint{1502.01428}].

\bibitem{Wei:2015ana}
  S.~W.~Wei, P.~Cheng and Y.~X.~Liu,
  [\eprint{1510.00085}].

\bibitem{Rajagopal:2014ewa}
  A.~Rajagopal, D.~Kubiz¨¾¨¢k and R.~B.~Mann,
  Phys.\ Lett.\ B {\bf 737}, 277 (2014)
  [\eprint{1408.1105}].

\bibitem{Hennigar:2015esa}
  R.~A.~Hennigar, W.~G.~Brenna and R.~B.~Mann,
  JHEP {\bf 1507}, 077 (2015)
    [\eprint{1505.05517}].



\bibitem{Frassino:2015oca}
  A.~M.~Frassino, R.~B.~Mann and J.~R.~Mureika,
 [\eprint{1509.05481}].

\bibitem{Oltean:2015lta}
  M.~Oltean, R.~J.~Epp, P.~L.~McGrath and R.~B.~Mann,
 [\eprint{1510.02858}].

\bibitem{Mo:2015xpa}
  J.~X.~Mo and W.~B.~Liu,
   [\eprint{1510.02625}].

\bibitem{Lee:2015wua}
  C.~O.~Lee,
     [\eprint{1510.06217}].

\bibitem{Xu:2013zea}
  W.~Xu, H.~Xu and L.~Zhao,
  Eur.\ Phys.\ J.\ C {\bf 74}, 2970 (2014)
  [\eprint{1311.3053}].

\bibitem{Kastor:2010gq}
  D.~Kastor, S.~Ray and J.~Traschen,
  Class.\ Quant.\ Grav.\  {\bf 27}, 235014 (2010)
  [\eprint{1005.5053}].

\bibitem{Xu:2014tja}
  H.~Xu, W.~Xu and L.~Zhao,
  Eur.\ Phys.\ J.\ C {\bf 74}, no. 9, 3074 (2014)
  [\eprint{1405.4143}].


\bibitem{Spallucci:2013osa}
 E.~Spallucci and A.~Smailagic,
  Phys.\ Lett.\ B {\bf 723}, 436 (2013)
  [\eprint{1305.3379}].

\bibitem{Spallucci:2013jja}
  E.~Spallucci and A.~Smailagic,
  J.\ Grav.\  {\bf 2013}, 525696 (2013)
    [\eprint{1310.2186}].

\bibitem{Belhaj:2014eha}
  A.~Belhaj, M.~Chabab, H.~El moumni, K.~Masmar and M.~B.~Sedra,
  Eur.\ Phys.\ J.\ C {\bf 75}, no. 2, 71 (2015)
  [\eprint{1412.2162}].

\bibitem{Zhang:2014fsa}
  L.~C.~Zhang, H.~H.~Zhao, R.~Zhao and M.~S.~Ma,
  Adv.\ High Energy Phys.\  {\bf 2014}, 816728 (2014).

\bibitem{Zhao:2014eja}
  J.~X.~Zhao, M.~S.~Ma, L.~C.~Zhang, H.~H.~Zhao and R.~Zhao,
  Astrophys.\ Space Sci.\  {\bf 352}, 763 (2014).

\bibitem{Lan:2015bia}
  S.~Q.~Lan, J.~X.~Mo and W.~B.~Liu,
  Eur.\ Phys.\ J.\ C {\bf 75}, no. 9, 419 (2015)
    [\eprint{gr-qc/1503.07658}].

\bibitem{Johnson:2013dka}
  C.~V.~Johnson,
  JHEP {\bf 1403}, 047 (2014)
    [\eprint{1306.4955}].

\bibitem{Caceres:2015vsa}
  E.~Caceres, P.~H.~Nguyen and J.~F.~Pedraza,
  JHEP {\bf 1509}, 184 (2015)
  [\eprint{1507.06069}].

\bibitem{Nguyen:2015wfa}
  P.~H.~Nguyen,
  JHEP {\bf 1512}, 139 (2015)
  [\eprint{gr-qc/1508.01955}].


\bibitem{Sun:2016til}
  Y.~Sun, H.~Xu and L.~Zhao,
 [\eprint{gr-qc/1606.06531}].




\bibitem{Boulware}
 D.~G.~Boulware and S.~Deser,
   Phys.\ Rev.\ Lett.\  {\bf 55}, 2656 (1985).

\bibitem{RGCai2002}
R.~-G.~Cai,
  Phys.\ Rev.\ D {\bf 65}, 084014 (2002)
    [\eprint{hep-th/0109133}].


\bibitem{Wiltshir}
D.~L.~Wiltshire,
   Phys.\ Lett.\ B {\bf 169}, 36 (1986).


\bibitem{Cvetic}
  M.~Cvetic, S.~'i.~Nojiri and S.~D.~Odintsov,
   Nucl.\ Phys.\ B {\bf 628}, 295 (2002)
    [\eprint{hep-th/0112045}].

\bibitem{Kofinas:2006hr}
  G.~Kofinas and R.~Olea,
  Phys.\ Rev.\ D {\bf 74}, 084035 (2006)
  [\eprint{hep-th/0606253}].


\bibitem{Kastor:2011qp}
  D.~Kastor, S.~Ray and J.~Traschen,
  Class.\ Quant.\ Grav.\  {\bf 28}, 195022 (2011)
  [\eprint{1106.2764}].


\bibitem{Dehghani:2005vh}
  M.~H.~Dehghani and M.~Shamirzaie,
    Phys.\ Rev.\ D {\bf 72}, 124015 (2005)
    [\eprint{hep-th/0506227}].

\bibitem{Kofinas:2007ns}
  G.~Kofinas and R.~Olea,
  JHEP {\bf 0711}, 069 (2007)
  [\eprint{0708.0782}].


\bibitem{Dehghani:2009zzb}
  M.~H.~Dehghani and R.~Pourhasan,
    Phys.\ Rev.\ D {\bf 79}, 064015 (2009)
     [\eprint{0903.4260}].

\bibitem{Zou:2010yr}
  D.~Zou, R.~Yue and Z.~Yang,
  Commun.\ Theor.\ Phys.\  {\bf 55}, 449 (2011)
  [\eprint{1011.2595}].

\bibitem{Banerjee:2010da}
  R.~Banerjee, S.~Ghosh and D.~Roychowdhury,
  Phys.\ Lett.\ B {\bf 696}, 156 (2011)
    [\eprint{1008.2644}].

\bibitem{Lala:2011np}
  A.~Lala and D.~Roychowdhury,
  Phys.\ Rev.\ D {\bf 86}, 084027 (2012)
  [\eprint{1111.5991}].

\bibitem{Zhao:2014sra}
  Z.~Zhao and J.~Jing,
  JHEP {\bf 1411}, 037 (2014)
  [\eprint{1405.2640}].

\bibitem{Mo:2013ela}
  J.~X.~Mo and W.~B.~Liu,
  Phys.\ Lett.\ B {\bf 727}, 336 (2013).

\bibitem{Mo:2014lza}
  J.~X.~Mo,
  Europhys.\ Lett.\  {\bf 105}, 20003 (2014).

\bibitem{Mo:2014wca}
  J.~X.~Mo, G.~Q.~Li and W.~B.~Liu,
  Phys.\ Lett.\ B {\bf 730}, 111 (2014).

\bibitem{Mo:2014mba}
  J.~X.~Mo and W.~B.~Liu,
  Phys.\ Rev.\ D {\bf 89}, no. 8, 084057 (2014)
  [\eprint{1404.3872}].






\end{thebibliography}

\providecommand{\href}[2]{#2}\begingroup
\footnotesize\itemsep=0pt
\providecommand{\eprint}[2][]{\href{http://arxiv.org/abs/#2}{arXiv:#2}}

\end{document}